\documentclass[12pt]{article}
\usepackage{latexsym}
\usepackage{epsfig,amssymb,euscript,slashed}
\usepackage{amsmath}
\usepackage{cite} 
\usepackage{array,calc,epsfig}
\usepackage{bbm}
\usepackage{fancybox}
\usepackage[linktocpage]{hyperref}

\def\be{\begin{equation}}
\def\ee{\end{equation}}
\def\bseq{\begin{subequations}}
\def\eseq{\end{subequations}}

\def\bea{\begin{eqnarray}}
\def\eea{\end{eqnarray}}

\def\bseq{\begin{subequations}}
\def\eseq{\end{subequations}}
\def\beq{\begin{equation}}
\def\eeq{\end{equation}}

\numberwithin{equation}{section} %%
\usepackage[]{graphicx}

\def\cald{{\cal D}}

\def\complex      {{\mathbb C}}

\def\zet          {{\mathbb Z}}

\def\sqr#1#2{{\vcenter{\vbox{\hrule height.#2pt
 \hbox{\vrule width.#2pt height#1pt \kern#1pt \vrule width.#2pt}\hrule
 height.#2pt}}}}

\def\slashchar#1{\setbox0=\hbox{$#1$}           % set a box for #1
\dimen0=\wd0                                 % and get its size
\setbox1=\hbox{/} \dimen1=\wd1               % get siste of /
\ifdim\dimen0>\dimen1                        % #1 is bigger
\rlap{\hbox to \dimen0{\hfil/\hfil}}      % so center / in box
#1                                        % and print #1
\else                                        % / is bigger
\rlap{\hbox to \dimen1{\hfil$#1$\hfil}}   % so center #1
/                                         % and print /
\fi}

\begin{document}
\font\cmss=cmss10 \font\cmsss=cmss10 at 7pt

\begin{flushright}{\scriptsize DFPD-13-TH-21 \\  \scriptsize QMUL-PH-13-12}
\end{flushright}
\hfill
\vspace{18pt}
\begin{center}
{\Large 
%\textbf{Microstates at the boundary of AdS:\\ non-linear solutions}
\textbf{Superdescendants of the D1D5 CFT \\ and their dual 3-charge geometries}
}
\end{center}

\vspace{8pt}
\begin{center}
{\textsl{ Stefano Giusto$^{\,a, b}$ and Rodolfo Russo$^{\,c, d}$}}

\vspace{1cm}

\textit{\small ${}^a$ Dipartimento di Fisica ed Astronomia ``Galileo Galilei",  Universit\`a di Padova,\\Via Marzolo 8, 35131 Padova, Italy} \\  \vspace{6pt}

\textit{\small ${}^b$ I.N.F.N. Sezione di Padova,
Via Marzolo 8, 35131 Padova, Italy}\\
\vspace{6pt}

\textit{\small ${}^c$ Centre for Research in String Theory, School of Physics and Astronomy\\
Queen Mary University of London,
Mile End Road, London, E1 4NS,
United Kingdom}\\
\vspace{6pt}

\textit{\small ${}^d$ Laboratoire de Physique Th\'eorique de L'Ecole Normale Sup\'erieure\\
24 rue Lhomond, 75231 Paris cedex, France}\\

\end{center}

\vspace{12pt}

\begin{center}
\textbf{Abstract}
\end{center}

\vspace{4pt} {\small
\noindent 
We describe how to obtain the gravity duals of semiclassical states in the D1-D5 CFT that are superdescendants of a class of RR ground states. On the gravity side, the configurations we construct are regular and asymptotically reproduce the 3-charge D1-D5-P black hole compactified on $S^1\times T^4$. The geometries depend trivially on the $T^4$ directions but non-trivially on the remaining 6D space. In the decoupling limit, they reduce to asymptotically AdS$_3 \times S^3 \times T^4$ spaces that are dual to CFT states obtained by acting with (exponentials of) the operators of the superconformal algebra. As explicit examples, we generalise the solution first constructed in arXiv:1306.1745 and discuss another class of states that have a more complicated dual geometry. By using the free orbifold description of the CFT we calculate the average values for momentum and the angular momenta of these configurations. Finally we compare the CFT results with those obtained in the bulk from the asymptotically $M^{1,4} \times S^1\times T^4$ region.}

\vspace{1cm}

%\noindent {\em Possible comment ..........................................................................................................................................}

\thispagestyle{empty}

\vfill
\vskip 5.mm
\hrule width 5.cm
\vskip 2.mm
{
\noindent  {\scriptsize e-mails:  {\tt stefano.giusto@pd.infn.it, r.russo@qmul.ac.uk} }
}

\newpage

\setcounter{footnote}{0}

\tableofcontents

%\newpage

 %%%%%%%%%%%%%%%%%%%%%%%%%%%%%%%%%%%%%%%%%

\section{Introduction}

Supergravity theories represent a perfect laboratory where to study in
a more tractable setup many conceptual issues related to black holes
physics. In particular supersymmetric theories in 5D played an
important role in the microscopic analysis of the Bekenstein-Hawking
entropy for extremal black
holes~\cite{Strominger:1996sh,Callan:1996dv}. In this paper we will
consider exactly this setup, where type IIB string theory is
compactified on $S^{1} \times T^4$: the 4D torus\footnote{We will
  focus on configurations that are independent of the compact 4D
  space, so our discussion applies equally well also to the case where
  the $T^4$ is substituted with a ``small'' K3.}  is of stringy size,
while the radius of the $S^1$ is much bigger than the string length $R
\gg \sqrt{\alpha'}$.

According to the ``fuzzball'' conjecture a black hole geometry (as a
solution of the classical supergravity equations of motion) is an
effective description of the gravitational physics that generically
breaks down at the scale of the black hole horizon~\cite{Mathur:2005zp,Bena:2007kg,Skenderis:2008qn,Mathur:2008nj,Balasubramanian:2008da,Chowdhury:2010ct,Mathur:2012zp,Mathur:2012dxa}. This
means that, even in the limit where the Planck length is set to zero
$\ell_P\to 0$ by keeping the horizon radius $R_S$ fixed, it should not
be possible to describe the dynamics of {\em all} types of light
probes close to the horizon by using Quantum Field Theory in the black
hole background. While it is difficult to prove this directly, in the
simplest cases it is possible to support this proposal with some very
explicit calculations. 

A popular approach has been to focus on the special subclass of the
black hole microstates that, from a quantum point of view, correspond
to semiclassical (i.e. coherent) states. It is natural to expect that
this type of states can be described by a classical geometry that
solves the type IIB supergravity equations. A first goal is to
construct these geometries and check that at spatial infinity they behave
as the black hole solutions, but typically start differing from it at scales
equal or larger than $R_S$. This problem has been solved in the
$1/4$-BPS case~\cite{Lunin:2001fv,Lunin:2001jy,Lunin:2002bj,Lunin:2002iz,Kanitscheider:2007wq}. However in this case the connection with
the standard black hole physics is less transparent, since according
to the usual Bekenstein's formula the corresponding black hole has
zero entropy and one is forced to go beyond the two derivative
approximation on the gravity side~\cite{Dabholkar:2004yr} (see
also~\cite{Sen:2009bm} for an analysis of this case in different
duality frames). The study of the $1/8$-BPS case, which corresponds to
a large black hole, has been successful in building many interesting
supergravity solutions~\cite{Lunin:2004uu,Giusto:2004id,Giusto:2004ip,Giusto:2004kj,Bena:2005va,Berglund:2005vb,Bena:2006is,Bena:2006kb,Ford:2006yb,Lunin:2012gp,Gibbons:2013tqa,Giusto:2013rxa} with the same asymptotic behaviour as
the Strominger-Vafa black hole, but so far it has not been possible to
really probe the whole phase space of the black
hole microstates.

The aim of this paper is to provide a constructive approach for
building new microstate geometries. As it will be clear later, our
technique applies only to a special type of semiclassical states and
so these solutions cannot account, even qualitatively, for the
dependence of the black hole entropy on the asymptotic
charges. However, the configurations we consider have some important
features that should be shared by the supergravity solutions
corresponding to generic semiclassical states. First, of course, these
geometries solve the full non-linear supergravity equations and do not
have horizons or naked singularities. Moreover they have an AdS throat
region while at the spatial infinity reduce to the standard 3-charge
Strominger-Vafa black hole. Each of these two regions is related to a
(different) microscopic CFT interpretation of the solutions.

If we focus on the core of each solution (i.e. we consider the ``inner
region'' according to the nomenclature of~\cite{Mathur:2011gz}), then
we have an asymptotically AdS$_3 \times S^3 \times T^4$ $1/8$-BPS
geometry.  By the AdS/CFT duality, this geometry should correspond to
the backreaction of a particular state in the dual CFT. In our case
the dual CFT is a (deformation of a) $1+1$ dimensional sigma-model
with target space $(T^4)^N/S_N$, where $N$ is the product of the D1
and D5 charges of the black hole $N=n_1 n_5$. This CFT has $(4,4)$
supersymmetry and the supercharges transform under a $SU(2)\times
SU(2)$ R-symmetry. As done in~\cite{Mathur:2011gz}, we consider states
that are super-descendants of Ramond-Ramond (RR) ground states. On the
gravity side, the ground states are dual to the 2-charge geometries
and the action of the Virasoro or the affine R-symmetry generators
correspond to changes of coordinates that do not vanish at the
boundary of AdS.  We construct in this way asymptotically AdS 3-charge
gravity solutions that have a well-understood CFT dual, corresponding
to a descendant of a RR ground state. The geometries we obtain
explicitly depend on the null-like coordinate $v=(t+y)/\sqrt{2}$,
where $y$ is the coordinates along the ``large'' $S^1$. We identify
the CFT origin of this $v$-dependence: $v$-dependent geometries are
dual to CFT states that are not eigenstates of the momentum operator
$L_0-\tilde L_0$.  In this respect the geometries we consider are
qualitatively different from previously known examples of 3-charge
geometries with well-understood CFT duals
\cite{Giusto:2004id,Giusto:2004ip}: those geometries did not depend on
$v$, a reflection of the fact that the dual states where exact
eigenstates of momentum.  We expect the supergravity solutions
corresponding to generic 3-charge microstates to lie in the class of
$v$-dependent geometries.\footnote{The same conclusion was reached
  from a different perspective in~\cite{Bena:2011uw}. For attempts to
  construct $v$-dependent solutions representing unbound
  superpositions of D1 and D5 charges carrying momentum,
  see~\cite{Bena:2011dd,Niehoff:2012wu,Vasilakis:2013tjs}.}

As a further step, we discuss the generalisation of the backgrounds
obtained in this way to asymptotically flat solutions with the same
charges as the Strominger-Vafa black hole. The asymptotically flat
part is directly related to the description of the microstate in terms
of D1 and D5-branes. As shown
in~\cite{Giusto:2009qq,Black:2010uq,Giusto:2011fy,Giusto:2012jx} it is
possible to use the worldsheet CFT describing the open strings
stretched between the D-branes to derive the long distance behaviour of
the various supergravity fields, including the multipole terms which
are absent in the black hole geometry. In this paper we will not
pursue this point of view at the quantitative level. However, we will
use it as a general guiding principle by requiring that the
asymptotically flat extension falls off at infinity in a way that is
compatible with the worldsheet CFT. So it is possible that the
solutions constructed in this paper fall in the class discussed
in~\cite{Giusto:2012jx} and are related to specific D-brane
configurations.

The extension to an asymptotically flat configuration is in general a
non-trivial task, especially for $v$-dependent solutions.
In this paper we focus on a restricted class of geometries, 
for which the metric in the four spatial non-compact dimensions is
$v$-independent up to a conformal factor, at least in an appropriate system
of coordinates.\footnote{In the example of Section \ref{gvd} the coordinate system
in which the 4D part of the metric is conformally $v$-independent does not coincide
with the coordinate system in which the asymptotic limit of the geometry looks
explicitly like $M^{1,4} \times S^1\times T^4$.} These geometries are 
descendants of particular 2-charge seed solutions, those that
are associated with Lunin-Mathur profiles \cite{Lunin:2001fv} whose
projection on $\mathbb{R}^4$ is a circle. Notice that this does not automatically
imply that the seed solution has a $U(1)\times U(1)$ symmetry, as one could 
admit configurations with non-trivial density profiles, as the one in Section \ref{2cfullg} or the ones 
discussed in~\cite{Bena:2010gg}. For this class of seed solutions, it is
possible to study the problem of building the geometry corresponding
to the descendant states from the $1+3$ dimensional point of view
discussed in~\cite{Niehoff:2013kia}. This formalism allows to extend
most of the metric components and of the fluxes in a very
straightforward and algebraic way. Only the components of the metric that
are associated with the angular momentum have to be found by solving a system of partial differential
equations. In order to smoothly connect the asymptotically flat
part with the inner region one has to modify the solution at the
centre of AdS by corrections of the
order of $R_{\rm AdS}/R \ll 1$. We show how this can be done in some
specific example.

Our paper builds on several previous works focusing on the
construction of 3-charge geometries. The idea of exploiting the
geometric version of the algebra generators is presented
in~\cite{Mathur:2003hj}, where it was used to derive a linearised
solution in the context of 6D supergravity. Always at the linearised
level, \cite{Mathur:2011gz} presents a systematic study of all algebra
generators, while~\cite{Mathur:2012tj,Lunin:2012gp} focuse on the generators related to the $T^4$
part. Here we will focus on the
generators that act non-trivially on the $\mathbb{R}^4$ part and thus change
more substantially the 2-charge geometry taken as the starting point.
The example presented in Section \ref{gvind} 
was briefly considered in~\cite{Giusto:2013rxa}, as a
possible 10D uplift of the linearised 6D
solution of~\cite{Mathur:2003hj}. In this paper we provide more details
about this construction and generalise it to arbitrary values of the
rotation parameter. We also discuss the corresponding state in the
dual CFT and provide some basic checks about the validity of this identification. The 10D uplift of the solution of~\cite{Mathur:2003hj} is not unique and another possibility appears as a particular case of the linearised solutions constructed in~\cite{Shigemori:2013lta}. That paper considers, at the perturbative level and restricting to the inner region, the action of an SU(2) generator at level zero in the NS sector on a generic 2-charge Lunin-Mathur geometry and shows that the transformed metrics generically have a $v$-dependent 4D part. 

It is also possible to act with algebra generators of level $n>0$: in Section \ref{gvd} we construct an example of a geometry generated in this way and Section \ref{svd} discusses the corresponding microscopic description. As in~\cite{Shigemori:2013lta}, the 4D part of the transformed metric is $v$-dependent; however the 4D metric can be made (conformally) $v$-independent by going to an appropriate system of coordinates, at the price of having an asymptotic limit which is not explicitly flat. We exploit this fact to  extend the geometry, at the non-linear level, to the asymptotic region: we discuss both the regularity conditions in the core and the asymptotically flat behaviour, and compare the charges computed in the CFT and the gravitational descriptions. In this case the comparison between the gravity and the dual CFT description is less direct. In our explicit example, we focus on a configuration that is an eigenstate of the angular momentum operators, so the corresponding values are quantised and agree in a straightforward way on the supergravity and the dual CFT side. However, as mentioned above, the microstate under analysis is not an eigenvector of the momentum operator and so for this observable we can read only an average value that depends on the (continuous) parameter defining the coherent state. The identification of this parameter in the gravity and the microscopic descriptions is unambiguous only in the decoupling limit (where the CFT conformal superalgebra can be realised geometrically). We propose that the two descriptions match also at finite values of AdS radius only after including corrections of order $R_{\rm AdS}/R$ in the dictionary between the parameter defining the coeherent state and that defining the asymptotically flat geometry.

We briefly outline the plan of the paper. The supergravity equations that must be satisfied by 1/8 BPS solutions are reviewed in Section~\ref{section:sugra}. Section~\ref{section:2charge} introduces the 2-charge seed solutions that form the starting point of our construction and the coordinate transformations that are used to add momentum, and describes their interpretation in the dual CFT. The general solution generating technique that we employ to extend the geometries to the asymptotically flat region is outlined in Section~\ref{section:generating} and is applied in Section~\ref{section:examples} to the construction of two different 3-charge microstates. The charges of the states dual to these geometries are computed on the CFT side in Section~\ref{section:CFTcharges}
and are compared with the ones extracted from the asymptotic region of the geometry in Section~\ref{section:CFTmatching}. In the discussion section we summarize the main qualitative features of our solutions and the general conclusions on the structure of black hole microstate geometries that we think could be drawn from them. The generalization to generic rotation parameter of the solution of Section~\ref{gvind} is detailed in the Appendix.

\section{Supergravity equations}\label{section:sugra}
The general solution of type IIB supergravity compactified on $T^4\times S^1$ preserving the same supersymmetries as the D1-D5-P system was found in~\cite{Giusto:2013rxa}, under the sole assumption that the geometry is invariant under rotations of $T^4$. The solution can be written as
\begin{subequations}\label{ansatzsummary}
\allowdisplaybreaks
 \begin{align}
d s^2_{(10)} &= -\frac{2\alpha}{\sqrt{Z_1 Z_2}}\,(d v+\beta)\,\Big[d u+\omega + \frac{\mathcal{F}}{2}(d v+\beta)\Big]+\sqrt{Z_1 Z_2}\,d s^2_4+\sqrt{\frac{Z_1}{Z_2}}\,d \hat{s}^2_{4}\, ,\\
e^{2\phi}&=\alpha\,\frac{Z_1}{Z_2}\, ,\\
B&= -\frac{\alpha\,Z_4}{Z_1 Z_2}\,(d u+\omega) \wedge(d v+\beta)+ a_4 \wedge  (d v+\beta) + \delta_2\,,\\ 
 C_0&=\frac{Z_4}{Z_1}\, ,\\
C_2 &= -\frac{\alpha}{Z_1}\,(d u+\omega) \wedge(d v+\beta)+ a_1 \wedge  (d v+\beta) + \gamma_2\,,\\ 
C_4 &= \frac{Z_4}{Z_2}\, \hat{\mathrm{vol}}_{4} - \frac{\alpha\,Z_4}{Z_1 Z_2}\,\gamma_2\wedge (d u+\omega) \wedge(d v+\beta)+x_3\wedge(d v + \beta)\,,
\end{align}
\end{subequations}
where
\be
\alpha = \frac{Z_1 Z_2}{Z_1 Z_2 - Z_4^2}\,.
\ee
The 10D space-time is split into the compact manifold $T^4$, endowed with a flat metric $d \hat{s}^2_{4}$, the four non-compact spatial directions, diffeomorphic to $\mathbb{R}^4$, over which we define a generically non-trivial Euclidean metric $d s^2_{4}$, and the time and $S^1$ directions, $t$ and $y$, that we parametrize with light-cone coordinates 
\be
u=\frac{t-y}{\sqrt{2}}\,,\quad v=\frac{t+y}{\sqrt{2}}\,.
\ee
The remaining ingredients defining the ansatz are: the 0-forms on $\mathbb{R}^4$ $Z_1$, $Z_2$, $Z_4$ and $\mathcal{F}$; the 1-forms $\beta$, $\omega$, $a_1$ and $a_4$;
the 2-forms  $\gamma_2$ and $\delta_2$; the 3-form $x_3$. One can also introduce a 1-form $a_2$ and a 2-form $\gamma_1$ that appear in $C_6$, the 6-form dual to $C_2$, in a way analogous to how $a_1$ and $\gamma_2$ appear in $C_2$. All these objects, including  $d s^2_{4}$, 
depend in general on the coordinate $v$ and the $\mathbb{R}^4$ coordinates $x^i$.
The constraints that these geometric data have to satisfy in order to preserve supersymmetry and satisfy the equations of motion have been derived in~\cite{Giusto:2013rxa}. As explained there,  generalizing~\cite{Bena:2011dd}, the subset of these constraints which is intrinsically non-linear, and hence  hardest to solve, involves the 4D metric 
$d s^2_{4}$ and the 1-form $\beta$:
\begin{subequations}\label{eqJbeta}
 \begin{align}\label{eqJ}
& d J_A= \frac{d}{d v}(\beta\wedge J_A)\,,\quad *_4 J_A = - J_A\,,\quad J_A\wedge J_B = -2 \,\delta_{AB}\,\mathrm{vol}_4\,, \\
&\label{eqbeta} *_4\cald\beta=\cald\beta\,,
\end{align}
\end{subequations}
where 
\be
\cald \equiv  d - \beta \wedge \frac{d}{d v}
\ee
with $d$ the differential on $\mathbb{R}^4$, 
$*_4$ and $\mathrm{vol}_4$ denote the Hodge dual and the volume form associated with $d s^2_{4}$ and $J_A$, $A=1,2,3$, are 2-forms defining an almost complex structure for $d s^2_{4}$. 
The problem simplifies if one assumes $d s^2_{4}$ and $\beta$ to be $v$-independent: in this case 
the equations imply that $d s^2_{4}$ is hyperk\"ahler and that the $\beta$ field-strength, $d\beta$, is a self-dual 2-form on this hyperkahler space. It turns out that for the microstates we consider, coordinates can be chosen in such a way that $d s^2_{4}$ is simply the flat metric on $\mathbb{R}^4$ and $\beta$ does not depend on $v$. In the following we will thus restrict to the simplified class of solutions where
$d s^2_{4}$ and $\beta$ are $v$-independent but all the other geometric data are allowed to depend on $v$.  The equations these data have to satisfy are summarized below. 

\begin{itemize}

 \item Equations for $ Z_1, a_2, \gamma_1$:
 \begin{subequations}\label{eqZ1a2}
 \begin{align}\label{eqZ1}
& *_4 \cald Z_1 = \cald \gamma_1 - a_2\wedge d\beta\,,\\
&\label{eqTh2}
\Theta_2 = *_4 \Theta_2  \quad~~~~\mathrm{with}\quad \Theta_2 = \cald a_2 + \dot \gamma_1\,.
\end{align}
\end{subequations}

\item Equations for $Z_2, a_1, \gamma_2$:
 \begin{subequations}\label{eqZ2a1}
 \begin{align}\label{eqZ2}
& *_4\cald Z_2 = \cald \gamma_2 - a_1\wedge d\beta\,,\\
&\label{eqTh1}
\Theta_1= *_4 \Theta_1  \quad~~~~\mathrm{with}\quad \Theta_1 = \cald a_1 + \dot \gamma_2\,.
\end{align}
\end{subequations}

\item Equations for $Z_4, a_4, \delta_2$:
 \begin{subequations}\label{eqZ4a4}
 \begin{align}
& \label{eqZ4}
 *_4 \cald Z_4 = \cald \delta_2 - a_4\wedge d\beta\,,\\
&\label{eqTh4}
\Theta_4 = *_4 \Theta_4 \quad~~~~\mathrm{with}\quad \Theta_4 = \cald a_4 + \dot\delta_2\,.
\end{align}
\end{subequations}

\item Equations for $\omega,\mathcal{F}$:
\begin{subequations}\label{eqcalFomega}
\begin{align}\label{eqomega}
\cald \omega + *_4  \cald\omega + \mathcal{F} \,d\beta &= Z_1\, \Theta_1+ Z_2\,\Theta_2 -2\,Z_4\, \Theta_4 \,,\\
\label{eqcalF} *_4\cald *_4 \Bigl(\dot{\omega} -\frac{1}{2}\,\cald \mathcal{F}\Bigr) &=\dot{Z}_1\dot{Z}_2+Z_1 \ddot{Z}_2 + Z_2 \ddot{Z}_1 -(\dot{Z}_4)^2 -2 Z_4 \ddot{Z}_4
\nonumber\\
&-\frac{1}{2} *_4 \Big[\Theta_1\wedge \Theta_2 - \Theta_4 \wedge \Theta_4\Bigr]\,,
\end{align}
\end{subequations}

\item Equation for $x_3$:
\be\label{eqx3}
\cald x_3 - \Theta_4\wedge \gamma_2+ a_1 \wedge (\cald \delta_2- a_4\wedge d \beta) = Z_2^2\,*_4 \frac{d}{d v}\Bigl(\frac{Z_4}{Z_2}\Bigr)\,.
\ee

\end{itemize}

Above we gave the equations for the gauge potentials: $\gamma_1$ and $a_2$ are by themselves not gauge invariant, but the combination $\Theta_2$ is (and analogously for $\gamma_2$, $a_1$ and $\Theta_1$ and for   $\delta_2$, $a_4$ and $\Theta_4$).  It might  be useful to use also the gauge invariant form for the first three sets of the equations above:

\begin{itemize}

\item Equations for $ Z_1, \Theta_2$:
\begin{subequations}\label{eqZ1Th2}
 \begin{align}
& \label{eqZ1bis}
 \cald *_4 \cald Z_1 = - \Theta_2\wedge d\beta\,,\\
&\label{eqTh2bis}
\cald \Theta_2 = *_4 \cald \dot{Z_1}\,,\quad \Theta_1 = *_4 \Theta_1\,.
\end{align}
\end{subequations}

\item Equations for $ Z_2, \Theta_1$:
\begin{subequations}\label{eqZ2Th1}
 \begin{align}
& \label{eqZ2bis}
 \cald *_4 \cald Z_2 = - \Theta_1\wedge d\beta\,,\\
&\label{eqTh1bis}
\cald \Theta_1 = *_4 \cald \dot{Z_2}\,,\quad \Theta_2 = *_4 \Theta_2\,.
\end{align}
\end{subequations}

\item Equations for $ Z_4, \Theta_4$:
\begin{subequations}\label{eqZ4Th4}
 \begin{align}
& \label{eqZ4bis}
 \cald *_4 \cald Z_4 = - \Theta_4\wedge d\beta\,,\\
&\label{eqTh4bis}
\cald \Theta_4 = *_4 \cald \dot{Z_4}\,,\quad \Theta_4 = *_4 \Theta_4\,.
\end{align}
\end{subequations}

\end{itemize}

It is important to observe that each of the three subsets of equations for $Z_I, \Theta_J$ constitutes a linear system of differential equations in its respective unkwons. Moreover, once $Z_I$ and  $\Theta_J$ have been computed, the r.h.s. of eqs. (\ref{eqomega}), (\ref{eqcalF}) are completely specified, and the problem of finding $\omega$ and $\mathcal{F}$ also reduces to a linear one.  

\section{A particular class of 2-charge states}\label{section:2charge}

The simplest 2-charge solution of the equations summarised in the previous section is the naive superposition of D1 and D5-branes, which corresponds to setting all functions to zero, except $Z_1$ and $Z_2$ that should be harmonic function on $\mathbb{R}^4$. It is possible to introduce another harmonic function for ${\cal F}$ and obtain the simplest 3-charge solution by defining
\begin{equation}
  \label{eq:D1D5P}
  Z_1 = 1 + \frac{Q_1}{r^2}~,~~~
  Z_2 = 1 + \frac{Q_5}{r^2}~,~~~
  \frac{\cal F}{2} = - \frac{Q_p}{r^2}~
\end{equation}
and setting all other functions in~\eqref{ansatzsummary} to zero; as usual, the charges $Q_I$ must be integer multiples of the elementary D1, D5 and Kaluza-Klein charges
\begin{equation}
  \label{eq:Q1n1}
  Q_1 = \frac{(2\pi)^4 n_1 g_s (\alpha')^3}{V_4}~,~~~
  Q_5 =  g_s n_5 \alpha'~,~~~
  Q_p = \frac{(2\pi)^4 n_p g_s^2 (\alpha')^4}{V_4 R^2}~,
\end{equation}
where $g_s$ is the string coupling, $V_4$ the volume of the $T^4$, $R$ the radius of the $S^1$ and $\alpha'$ the Regge slope. In this paper we are interested in finding less trivial solutions that correspond to {\em bound states} of D-branes. We will first introduce a class of 2-charge configurations (i.e. solutions with ${\cal F}=0$) and discuss them both from the bulk and the dual CFT point of view. Then in the next section we will describe a constructive technique that allows to switch on a non-trivial momentum charge and construct a particular class of 3-charge geometries.

\subsection{The full geometry}\label{2cfullg}

All the 2-harge solutions corresponding to a D1-D5 bound state were constructed in~\cite{Lunin:2001fv,Lunin:2002iz,Kanitscheider:2007wq} by going to a duality frame where the system is described in terms of a fundamental string with a pulse (the F1-P frame). In this case the corresponding supergravity geometries are parametrised by a curve $g_A(v)$ in $\mathbb{R}^4\times T^4$ describing the profile of the string. After applying a duality transformation on the known solution in the F1-P frame~\cite{Callan:1995hn,Dabholkar:1995nc}, it was possible to write the solution for the D1-D5 configuration in terms of ``auxiliary'' profiles $g_A(v')$, that do not have any direct geometric meaning in the new duality frame. As already said, in this paper we will focus on the subclass of D1-D5 solutions invariant under the rotations of the $T^4$. In this case the most general 2-charge configuration is given in terms of five functions: four $g_i(v')$ corresponding to the F1 profile in $\mathbb{R}^4$ and one extra function, here denoted as $g(v')$, describing the F1 profile in a particular direction of $T^4$ that plays a special role in the chain of dualities relating the F1-P and D1-D5 frame. This class of 2-charge solutions can be written in terms of the ansatz~\eqref{ansatzsummary} by choosing $ds_4^2$ to be the Euclidean flat metric and
\begin{subequations}\label{generaltwocharge}
\begin{align}
& Z_2 = 1 + \frac{Q_5}{L} \int_0^{L} \frac{1}{|x_i -g_i(v')|^2}\, dv'~,~~~
  Z_4 = - \frac{Q_5}{L} \int_0^{L} \frac{\dot{g}(v')}{|x_i -g_i(v')|^2} \, dv' \,,\\\label{Z1profile}
& Z_1 = 1 + \frac{Q_5}{L} \int_0^{L} \frac{|\dot{g}_i(v')|^2+|\dot{g}(v')|^2}{|x_i -g_i(v')|^2} \, dv' ~, ~~~ d\gamma_2 = *_4 d Z_2~,~~~d\delta_2 = *_4 d Z_4~,\\
& A = - \frac{Q_5}{L} \int_0^{L} \frac{\dot{g}_j(v')\,dx^j}{|x_i -g_i(v')|^2} \, dv' ~, ~~~ dB = - *_4 dA~, \\
& \beta = \frac{-A+B}{\sqrt{2}}~,~~~\omega = \frac{-A-B}{\sqrt{2}}~,~~~{\cal F}=0~,~~~a_1=a_4=x_3=0~,
\end{align}
\end{subequations}
where the dot on the profile functions indicates a derivative with respect to $v'$.

The simplest 2-charge solution\footnote{This solution was first found in \cite{Maldacena:2000dr,Balasubramanian:2000rt}.} can be obtained from the general solution~\eqref{generaltwocharge} by using a circular profile in the plane $x_{1,2}$
\be\label{profile0}
g_1(v') = a \cos\left( \frac{2\pi\,v'}{L}\right) \,,\quad 
g_2(v') = a \sin\left( \frac{2\pi\,v'}{L}\right) \,,\quad 
\ee 
with all other $g_A(v')$ components trivial.\footnote{$L$ represents the length of the multiply wound fundamental string that is dual to the D1-D5 system and is given by
\be
L= 2\pi \frac{Q_5}{R}\,.
\ee
}
In order to calculate the integrals over $v'$ it is useful to introduce the coordinates~\cite{Lunin:2001fv} $r,\theta, \phi, \psi$ as follows: if $\tilde r, \tilde \theta, \phi, \psi$ denote a set of coordinates for $\mathbb{R}^4$ defined as
\be\label{eq:z1z2}
z_1=x_1+i\, x_2 = \tilde r \,\sin\tilde \theta\,e^{i\,\phi}\,,\quad z_2=x_3+i\, x_4 = \tilde r \,\cos\tilde \theta\,e^{i\,\psi}\,,
\ee
then $r$ and $\theta$ are
\be\label{tttheta}
\tilde r^2 = r^2 + a^2 \sin^2\theta\,,\quad \cos^2\tilde\theta = \frac{r^2 \,\cos^2\theta}{r^2+a^2 \sin^2\theta}\,.
\ee 
Then we have
\begin{subequations}\label{luninmathurtwocharge}
\allowdisplaybreaks
\begin{align}
d s^2_4 &= (r^2+a^2 \cos^2\theta)\Bigl(\frac{d r^2}{r^2+a^2}+ d\theta^2\Bigr)+(r^2+a^2)\sin^2\theta\,d\phi^2+r^2 \cos^2\theta\,d\psi^2\,,\label{ds4flat}\\
\beta &=  \frac{R\,a^2}{\sqrt{2}\,(r^2+a^2 \cos^2\theta)}\,(\sin^2\theta\, d\phi - \cos^2\theta\,d\psi)\,,\\
Z_1 &= 1+\frac{Q_1}{r^2+a^2 \cos^2\theta}\,,\\
Z_2 &=  1+\frac{Q_5}{r^2+a^2 \cos^2\theta}\,,\quad a_1=0\,,\quad \gamma_2 = -Q_5\,\frac{(r^2+a^2)\,\cos^2\theta}{r^2+a^2\cos^2\theta}\,d\phi\wedge d\psi\,,\\
Z_4 &=  0\,,\quad a_4 =0\,,\quad \delta_2=0\,,\\
\omega &=  \frac{R\,a^2}{\sqrt{2}\,(r^2+a^2 \cos^2\theta)}\,(\sin^2\theta\,d\phi + \cos^2\theta\,d\psi)\,, \\
\mathcal{F} &=  0\,.
\end{align}
\end{subequations}
Note that the 4D metric $d s^2_4$ in (\ref{ds4flat}) is just flat $\mathbb{R}^4$ written in non-standard coordinates. From the expression of $Z_1$ in terms of the profile (\ref{Z1profile}), it is easy to derive the relation between the radius of the ``large'' $S^1$, the charges $Q_I$ and the parameter $a$:
\be\label{Q1Q5Ra}
R = \frac{\sqrt{Q_1 Q_5}}{a}\,.
\ee

In this paper we will focus on the 2-charge configurations that have a circular profile in $\mathbb{R}^4$, but can have a non-trivial $g(v')$ component or also a more complicated parametrisation than the one in~\eqref{profile0}; our goal is to use these 2-charge configurations as seeds for generating new 3-charge solutions. In priciple we could use any profile function whose 4D part $g_i$ stays in a plane. Another example of a 2-charge configuration in this class was discussed in~\cite{Kanitscheider:2007wq} and can be obtained by adding a non-trivial $g(v')$ to the Lunin-Mathur case~\eqref{profile0}
\be\label{profile}
g_1(v') = a \cos\left( \frac{2\pi\,v'}{L}\right) \,,\quad 
g_2(v') = a \sin\left( \frac{2\pi\,v'}{L}\right) \,,\quad 
g(v') = - b \sin \left(\frac{2\pi\,v'}{L}\right)\,,
\ee 
with all other components trivial. This choice yields a geometry that can be embedded in the ansatz~\eqref{ansatzsummary} as follows
\begin{subequations}\label{skenderistwocharge}
\begin{align}
d s^2_4 &= (r^2+a^2 \cos^2\theta)\Bigl(\frac{d r^2}{r^2+a^2}+ d\theta^2\Bigr)+(r^2+a^2)\sin^2\theta\,d\phi^2+r^2 \cos^2\theta\,d\psi^2\,,\label{ds4flat2}\\
\beta &=  \frac{R\,a^2}{\sqrt{2}\,(r^2+a^2 \cos^2\theta)}\,(\sin^2\theta\, d\phi - \cos^2\theta\,d\psi)\,,\\
Z_1 &= 1+\frac{R^2}{Q_5} \frac{a^2+\frac{b^2}{2}}{r^2+a^2 \cos^2\theta}+\frac{R^2\, a^2\, b^2}{2\,Q_5}\,\frac{\cos2\phi\,\sin^2\theta}{(r^2+a^2 \cos^2\theta)(r^2+a^2)}\,,\\
Z_2 &=  1+\frac{Q_5}{r^2+a^2 \cos^2\theta}\,,\quad a_1=0\,,\quad \gamma_2 = -Q_5\,\frac{(r^2+a^2)\,\cos^2\theta}{r^2+a^2\cos^2\theta}\,d\phi\wedge d\psi\,,\\
Z_4 &=  R\, a\, b\,\frac{\cos\phi\,\sin\theta}{\sqrt{r^2+a^2}\,(r^2+a^2 \cos^2\theta)}\,,\quad a_4 =0\,,\\
\delta_2 &= \frac{ -R\, a\, b\ \sin\theta}{\sqrt{r^2+a^2}}\,\Bigl[\frac{r^2+a^2}{r^2+a^2 \cos^2\theta}\cos^2\theta \cos\phi\, \,d\phi\wedge d\psi + \sin\phi\, \frac{\cos\theta}{\sin\theta}\,
d\theta\wedge d\psi \Bigr]\,,\\
\omega &=  \frac{R\,a^2}{\sqrt{2}\,(r^2+a^2 \cos^2\theta)}\,(\sin^2\theta\,d\phi + \cos^2\theta\,d\psi)\,, \\
\mathcal{F} &=  0\,.
\end{align}
\end{subequations}
Of course, by setting $b=0$ in~\eqref{skenderistwocharge} we recover the configuration in~\eqref{luninmathurtwocharge}. In presence of a non-zero $b$, also the relation between the D1 and D5 charges, the radius of the ``large'' $S^1$ and the parameters of the profile is modified and, instead of~\eqref{Q1Q5Ra}, we have
\be\label{Q1Q5Rab}
R = \sqrt{\frac{Q_1 Q_5}{a^2+\frac{b^2}{2}}}\,.
\ee

\subsection{The decoupling limit}\label{2cdl}

As usual the decoupling limit, sometimes denoted also as ``near-horizon'' limit,  is defined by cutting off the asymptotically flat part of the solution and focusing on the core of the geometry. In formulae we have
\begin{equation}
  \label{eq:declim}
  r\ll \sqrt{Q_i} \ll R~~(i=1,5)~.
\end{equation}
In the case of the 2-charge geometries discussed above this approximation amounts to neglecting the ``1'' in the warp factors $Z_1, Z_2$. For instace, the naive 2-charge geometry~\eqref{eq:D1D5P} with $Q_1,Q_5 \not=0$ and $Q_p=0$ reduces to AdS$_3 \times S^3 \times T^4$ in  Poincar\'e coordinates. In the case of the solution~\eqref{luninmathurtwocharge}, the geometry in the decoupling limit is just global AdS$_3 \times S^3 \times T^4$, as it can be made explicit by the coordinate redefinition
\be\label{spectralflow}
\phi \to\phi+ \frac{t}{R}\,,\quad \psi \to \psi + \frac{y}{R}\,.
\ee
Notice that this change of variables is non-trivial at the boundary of AdS and so it should have a meaning also on the CFT side: Eq.~\eqref{spectralflow} corresponds to a spectral flow of the dual CFT from the R to the NS sector in both the holomorphic and the anti-holomorphic sectors. This is consistent with the the fact that~\eqref{luninmathurtwocharge} corresponds to a ground state in the RR sector and is mapped to a particular chiral primary state -- the $SL(2,\mathbb{C})$ vacuum (i.e. global AdS) -- after the change of coordinates~\eqref{spectralflow}.

Let us review some other changes of coordinates that remain non-trivial at the boundary of the AdS region and have a dual CFT interpretation: on the bulk side we use them to set up our solution generating technique, while the corresponding CFT action is used to identify precisely the state dual to the new geometry obtained. All 2-charge solutions have a flat base metric $ds_4^2$ in the ansatz~\eqref{ansatzsummary} and it is useful to study the $SU(2)_L\times SU(2)_R$ isometries of this Euclidean space. They can be parametrised as follows:
\be\label{LRrot}
\left(\begin{array}{cc} z_1 & -\bar z_2\\ z_2 & \bar z_1 \end{array}\right)\to e^{-\frac{i}{2} \chi^{(j)}_L \sigma^j}\,\left(\begin{array}{cc} z_1 & -\bar z_2\\ z_2 & \bar z_1 \end{array}\right)\,e^{\frac{i}{2} \chi^{(j)}_R \sigma^j}\,,
\ee
with\footnote{What is relevant is the action on the coordinates at the boundary of the AdS geometry in the decoupling limit. Then we can neglect the difference between $\tilde\theta$ and $\theta$, see Eq.~\eqref{tttheta}, since in this limit $a\ll r$.}
\be
z_1 = \sin\theta\,e^{i\phi}\,,\quad z_2 = \cos\theta\,e^{i\psi}\,,
\ee
and $\sigma^i$ the usual Pauli matrices. The infinitesimal transformations corresponding to the left and right generators $L^j,R^j = \sigma^j/2$ read 
\begin{subequations} \label{eq:Lgen}
\begin{align}
  L^1 & = \frac{i}{2} \sin(\phi-\psi) \partial_\theta + \frac{i}{2} \cos(\phi-\psi) \cot\theta \partial_\phi + \frac{i}{2} \cos(\phi-\psi) \tan\theta \partial_\psi~, \\ 
  L^2 & = -\frac{i}{2} \cos(\phi-\psi) \partial_\theta + \frac{i}{2} \sin(\phi-\psi) \cot\theta \partial_\phi + \frac{i}{2} \sin(\phi-\psi) \tan\theta \partial_\psi ~, \\ \label{J3}
  L^3 & = -\frac{i}{2} (\partial_\phi - \partial_\psi)~,
\end{align}
\end{subequations}
and
\begin{subequations} \label{eq:Rgen}
\begin{align}
  R^1 & = \frac{i}{2} \sin(\phi+\psi) \partial_\theta + \frac{i}{2} \cos(\phi+\psi) \cot\theta \partial_\phi - \frac{i}{2} \cos(\phi+\psi) \tan\theta \partial_\psi~, \\ 
  R^2 & = -\frac{i}{2} \cos(\phi+\psi) \partial_\theta + \frac{i}{2} \sin(\phi+\psi) \cot\theta \partial_\phi - \frac{i}{2} \sin(\phi+\psi) \tan\theta \partial_\psi ~, \\ \label{tildeJ3}
  R^3 & = -\frac{i}{2} (\partial_\phi + \partial_\psi)~,
\end{align}
\end{subequations}
which satisfy the algebra $[L^j,L^k]=i \epsilon_{jkl} L^l$, $[L^j,R^k]=0$, and $[R^j,R^k]=i \epsilon_{jkl} R^l$. As an example, let us focus on the $R^j$'s sector: we can introduce the standard raising and lowering operators $R^\pm = R^1 \pm i R^2$
\begin{equation}
  \label{eq:Rpm}
  R^\pm = \frac{1}{2}  e^{\pm i(\phi+\psi)} \left( \pm \partial_\theta + i \cot\theta \partial_\phi -i \tan\theta \partial_\psi \right)~.
\end{equation}
Let us define the change of coordinates corresponding to a general spectral flow, which is generated by 
\begin{equation}
  \label{eq:gensf}
\Sigma = - \frac{t}{R} \partial_\phi - \frac{y}{R} \partial_\psi~.
\end{equation}
Notice that the transformation $e^{-\Sigma}$ corresponds to the flow from the R to the NS sector introduced in~\eqref{spectralflow}, while $e^{\Sigma}$ describes the inverse flow from the NS to R sector. Let us consider the following sequence of operations on a geometry corresponding to a R ground state: a flow from the R to NS, an action of the generators $R^\pm$, and finally an inverse flow back to the R sector. In formulae, we have
\begin{equation}
  \label{eq:RpmSigma}
  e^{\Sigma} R^\pm e^{-\Sigma} = e^{\mp i \frac{t+y}{R}} R^\pm = e^{\mp i \frac{\sqrt{2} v}{R}} R^\pm \equiv R^\pm_{\mp 1}~,
\end{equation}
where in the last step we identified the combined operation on the original configuration in the R sector as the action corresponding to an affine generator on the CFT side at level $\pm 1$. This is consistent with the standard transformations on the CFT side reviewed in the following subsection.

In the example of Section~\ref{gvind}, we will need also the finite transformation corresponding to the generator $R^2$. If $\theta'$, $\phi'$ and $\psi'$ are the coordinates after a finite rotation with a parameter $\chi^{(2)}_R$ in~\eqref{LRrot}, we have
\begin{subequations}\label{rotg}
\begin{align}\label{rot1}
\cos^2\theta' & = \cos^2\theta \cos^2\frac{\chi_R^{(2)}}{2} + \sin^2\theta\sin^2\frac{\chi_R^{(2)}}{2}-\frac{1}{2}\sin(2\theta) \cos(\phi+\psi) \sin\chi_R^{(2)}\,,\\ 
\tan \phi' & = \frac{\sin\theta\sin\phi\cos\frac{\chi_R^{(2)}}{2} - \cos\theta \sin\psi \sin\frac{\chi_R^{(2)}}{2}}{\sin\theta\cos\phi\cos\frac{\chi_R^{(2)}}{2} + \cos\theta \cos\psi \sin\frac{\chi_R^{(2)}}{2}}\,,\\
\tan \psi' & = \frac{\cos\theta\sin\psi\cos\frac{\chi_R^{(2)}}{2} + \sin\theta \sin\phi \sin\frac{\chi_R^{(2)}}{2}}{\cos\theta\cos\psi\cos\frac{\chi_R^{(2)}}{2} - \sin\theta \cos\phi \sin\frac{\chi_R^{(2)}}{2}}\,.\label{rot3}
\end{align}
\end{subequations}

\subsection{The dual CFT point of view} \label{dctpoint}

As mentioned in the introduction, type IIB string theory propagating in the geometries obtained in the decoupling limit is dual to a $(4,4)$ superconformal CFT, whose central charge $c$ is determined by the D1 and D5 charges~\eqref{eq:Q1n1}: $c=6 n_1 n_5 $. The simplest way to describe this CFT is to focus on the so-called orbifold point of its moduli space, where there is a free field representation\footnote{For the CFT description we follow the conventions of~\cite{Avery:2010qw}.} in terms of four bosonic fields $X^{\dot{A} A}(z,\bar{z})$ and four doublets of chiral/antichiral fermionic fields $\psi^{\alpha \dot{A}}(z),~\tilde\psi^{\dot{\alpha} \dot{A}}(\bar{z})$. The upper-case indices $A$ and $\dot{A}$ indicate the fundamental representation of the first and the second $SU(2)$ in the usual decomposition $SU(2)_1\times SU(2)_2$ of the $SO(4)_{T^4}$ acting on the coordinates of the $T^4$; similarly the Greek indices refer to the $SU(2)_L\times SU(2)_R$ arising from the decomposition of the $SO(4)_{\mathbb{R}^4}$ acting on $\mathbb{R}^4$. Since the bosonic fields transform as a vector of $SO(4)_{T^4}$, we can say that the target space of the CFT consists of $n_1 n_5$ copies of the compact space $T^4$, which accounts for the value of the central charge, modded out by the permutation group $S^{n_1 n_5}$. 

Actually the geometric description summarised in the previous section is not directly related to the CFT at the free orbifold point. In order to describe the gravity regime on the CFT one should switch on a vacuum expectation value for the twist fields that swap different copies of the $T^4$ target space \cite{Avery:2010er,Avery:2010hs}. In this paper we will stick to the free field description and consider states in the untwisted sector; when dealing with eigenstates of operators describing conserved charges, such as the angular momentum or the momentum charge, we assume that they are protected and do not change with the deformation parameters that move the CFT away from the orbifold point.

We are particularly interested in the $SU(2)_R$ R-symmetry algebra that can be identified with the geometric generators~\eqref{eq:Rgen}: in the CFT description this $SU(2)_R$ is realised by a current algebra satisfying the usual OPEs
\begin{subequations} \label{eq:curra}
  \begin{align}
    J^j(z) J^k(w) & \sim \frac{i \epsilon^{jkl}}{z-w} J^l(w) + \frac{c}{12} \frac{\delta^{jk}}{(z-w)^2}~, \\ 
    T(z) J^j(w) & \sim \frac{\partial J^j(w)}{z-w} + \frac{J^j(w)}{(z-w)^2}~, \\
    T(z) T(w)  & \sim \frac{c}{(z-w)^4} + \frac{2 T(w)}{(z-w)^2} + \frac{\partial T(w)}{z - w} ~,
  \end{align}
\end{subequations}
where $T(z)$ is the stress energy tensor. The commutation relations among the modes $T=\sum_n L_n z^{-n-2}$ and $J^i = \sum_n J^i_n z^{-n-1}$ are
\begin{subequations} \label{eq:virmod}
  \begin{gather}
   [L_m,L_n] = (m-n) L_{m+n} + \frac{c}{12} m(m^2-1) \delta_{m+n,0}~, \\
 [J^j_m,J^k_n] = i\epsilon^{jkl} J^l_{m+n} + \frac{c}{12} m \delta_{m+n,0}~,~~~~~
  [L_m,J^j_n]=-n J^j_{m+n}~.
 \end{gather}
\end{subequations}
At the orbifold point the untwisted sector is just the tensor product of $n_1 n_5$ free fields and we can introduce a subscript $\ell=1,\ldots,n_1 n_5$ labelling the various copies (that we will call strands) of the target space. Thus we have\footnote{By following the conventions of~\cite{Avery:2010qw}, we have $(\psi^{\alpha \dot{A}}_\ell)^\dagger = -\epsilon_{\alpha\beta} \epsilon_{\dot{A} \dot{B}} \psi^{\beta \dot{B}}_\ell$.}
\begin{subequations}
  \label{eq:Jfree}
  \begin{align}
  J^+ & \equiv J^1 + i J^2 = \sum_\ell \frac{1}{2} \psi^{1 \dot{A}}_\ell \epsilon_{\dot{A}\dot{B}} \psi^{1\dot{B}}_\ell =\sum_\ell \psi^{1\dot{1}}_\ell \psi^{1\dot{2}}_\ell \equiv \sum_\ell i \chi^1_\ell \chi^2_\ell~, \\
  J^- & \equiv J^1 - i J^2 = -\frac{1}{2} \sum_\ell \psi^{2 \dot{A}}_\ell \epsilon_{\dot{A}\dot{B}} \psi^{2\dot{B}}_\ell = - \sum_\ell \psi^{2\dot{1}}_\ell \psi^{2\dot{2}}_\ell \equiv \sum_\ell i \bar\chi^1_\ell \bar\chi^2_\ell~, \\
  J^3 & = -\sum_\ell \frac{1}{2} \psi^{1\dot{A}}_\ell \epsilon_{\dot{A}\dot{B}} \psi^{2\dot{B}}_\ell \equiv \sum_\ell \frac{1}{2}\left(\chi^1_\ell \bar\chi^1_\ell + \chi^2_\ell \bar\chi^2_\ell\right)~,
  \end{align}
\end{subequations}
where we introduced the standard complex fermions  $\psi^{1\dot{1}}_\ell \equiv i \chi^1_\ell$, $\psi_\ell^{1\dot{2}}\equiv \chi^2_\ell$.

As pointed out in~\cite{Schwimmer:1986mf}, the superconformal algebra can be realised in different inequivalent ways by allowing twisted boundary conditions on the supercurrents and the affine generators. In terms of the free fields this amounts to add an extra $e^{2\pi i \nu}$ to the monodromy of the $\bar\chi$'s as $z\to e^{2\pi i} z$ and an extra $e^{-2\pi i \nu}$ for the $\chi$'s. Thus in general the mode expansion of the fermions is
\begin{equation}
  \label{eq:chimode}
\chi = \sum_{\zet + 1/2} \chi_{r+\nu} z^{-r-\nu-\frac{1}{2}}~,~~~~
\bar\chi = \sum_{\zet + 1/2} \bar\chi_{r-\nu} z^{-r+\nu-\frac{1}{2}}~,
\end{equation}
where we understood all indices since these equations are valid for both $\chi^i$ and for each strand $\ell$. Then the currents satisfy $J^\pm(e^{2\pi i}z) =e^{\mp 2\pi i 2\nu} J^\pm(z)$, and in order to preserve the OPE's~\eqref{eq:curra}, it is necessary to deform the definition of $T(z)$ and $J^3(z)$:\begin{equation}
  \label{eq:TJnu}
  T_\nu(z) = T(z) - \frac{2 \nu}{z} J^3(z) + \frac{c \nu^2}{6 z^2}~,~~~
  J^3_\nu(z) = J^3(z) -\frac{c \nu}{6 z}~.
\end{equation}
When $\nu=0$ we are in the NS sector, while the case $\nu=-1/2$ describes the flow from NS to R boundary conditions discussed on the gravity side after Eq.~\eqref{eq:gensf}. In terms of modes this flow implies to following relations
\begin{subequations}
  \label{eq:NS2Rcft}
  \begin{align}    
 &  (J^\pm_R)_{n} = (J^\pm_{NS})_{n\pm 1}~,~~~
  (J^3_{R})_n = (J^3_{NS})_{n} + \frac{c}{12} \delta_{n,0}~,\\
 &  (L_R)_n = (L_{NS})_n + (J^3_{NS})_n + \frac{c}{24} \delta_{n,0}~.
  \end{align}
\end{subequations}

Let us now describe the dual CFT interpretation of the 2-charge solutions: in the decoupling limit, these configurations are dual to R ground states of the CFT briefly described above~\cite{Lunin:2001jy, Kanitscheider:2006zf}. If we start from the $SL(2,\complex)$ invariant vacuum $|0\rangle$ in the NS sector, the $\nu=-1/2$ flow yields the R ground state $|n_1 n_5/2 \rangle$ that is an eigenstate of $(J^3_R)_0$ (and $(\tilde{J}^3_R)_0$) with eigenvalue $n_1 n_5/2$, as it can be easily checked by using Eqs.~\eqref{eq:NS2Rcft}. At the orbifold point we can characterise this state in terms of the oscillators introduced in~\eqref{eq:chimode}: as usual all positive modes annihilate any ground state and, for the zero-modes, we have
\begin{equation}
  \label{eq:Rhws}
  \bar\chi_0^i |n_1 n_5/2\rangle \not=  0 ~,~~~
  \chi_0^i |n_1 n_5/2\rangle = 0\quad\mathrm{for}\quad i=1,2~.
\end{equation}
Moving to the gravity description, as usual we identify the $SL(2,\mathbb{C})$ vacuum state with the global AdS$_3 \times S^3 \times T^4$ geometry. As seen in the previous subsection, the ``geometric'' version of the spectral flow relates the AdS$_3 \times S^3$ to the decoupling limit of the solution~\eqref{luninmathurtwocharge}. Thus, as a first entry of the dictionary between CFT states and geometries, it is natural to identify the configuration defined by the circular profile~\eqref{profile0} and the CFT state $|n_1 n_5/2\rangle$, as they are both related to the vacuum by spectral flow.

The general mapping between R ground states and 2-charge geometries is reviewed in~\cite{Skenderis:2008qn}. For our purposes we will need to discuss just the state dual to the configuration~\eqref{skenderistwocharge} which is determined by the profile~\eqref{profile}. In terms of the free field discussed above the different R ground states are obtained by acting on $|n_1 n_5/2\rangle$ with $(\bar\chi^i_0)_\ell$ and $(\tilde{\bar\chi}^i_0)_\ell$. We will focus on the operator\footnote{The normalization has been chosen so as to have $\langle n_1 n_5| O^\dagger O|n_1 n_5 \rangle=\sum_\ell 1 = n_1 n_5$.}
\begin{equation}
  \label{eq:Osk}
  O = \frac{1}{\sqrt{2}} \sum_{\ell=1}^{n_1 n_5} \left[(\bar\chi^1_0)_\ell (\tilde{\bar\chi}^{2}_0)_\ell - (\bar\chi^2_0)_\ell (\tilde{\bar\chi}^{1}_0)_\ell \right] = -\frac{i}{\sqrt{2}} \sum_{\ell=1}^{n_1 n_5} \epsilon_{\dot{A}\dot{B}} (\psi^{2\dot{A}}_0)_\ell (\tilde{\psi}^{\dot{2}\dot{B}}_0)_\ell \equiv \sum_\ell O_\ell~.
\end{equation}
It commutes with any permutation of the strands and so it creates physical states when acting on the highest weight state $|n_1 n_5/2\rangle$. Notice that $O$ is also a scalar under the rotations $SO(4)_{T^4}$, since the indices $\dot{A},\;\dot{B}$ are saturated in an invariant way, and that its action decreases the eigenvalue of both $J^3_0$ and $\tilde{J}^3_0$ by $1/2$. The state dual to the profile~\eqref{profile} is~\cite{Kanitscheider:2007wq}
\begin{equation}
  \label{eq:Esk}
  |\psi_0\rangle = \sum_{k=0}^{n_1 n_5} \frac{C_k}{k!} O^k |n_1 n_5/2\rangle~,
\end{equation}
where the normalization $C_k$ is
\begin{equation}
  \label{eq:Ck}
 C_k \equiv \sqrt{\left(
    \begin{array}{c}
      n_1 n_5 \\ k
    \end{array}\right)
} A^{n_1 n_5-k} B^k
~.
\end{equation}
The parameters $A\,,B$ in the equation above are related to the amplitudes $a\,,b$ that characterise the profile~\eqref{profile}
\be\label{AandB}
A= \frac{1}{\sqrt{1+\eta^2}}\,,\quad B = - \frac{\eta}{\sqrt{1+\eta^2}}\,,
\ee
where
\be\label{etadef}
\eta = \frac{b}{\sqrt{2}\,a}\,.
\ee
Finally notice that the relation $A^2+B^2=1$ implies that $|\psi_0\rangle$ has unit norm. 

Even if it is still in the untwisted sector, this state is closer to a generic semiclassical configuration than the highest weight state: for instance, its strands are not equal and it is not an eigenstate of $J^3_0$ nor $\tilde{J}^3_0$; this is true even in the large $n_1 n_5$ limit, where the sum in~\eqref{eq:Esk} is peaked around $k_p \sim n_1 n_5 B^2$ with a width of order $2 A B\sqrt{n_1 n_5}$. Thus $|\psi_0\rangle$ is the linear combination of states with different eigenvalues of ${J}^3_0+\tilde{J}^3_0$, but the same (zero) eigenvalue of ${J}^3_0-\tilde{J}^3_0$. Remembering the gravity realization of the operators ${J}^3_0$ and $\tilde{J}^3_0$ given in (\ref{J3}) and (\ref{tildeJ3}), one sees that this matches the generic features of the supergravity solution~\eqref{skenderistwocharge} which has only a $U(1)$ invariance (corresponding to the shifts of the angle $\psi$) but depends explicitly on $\phi$ (contrary to what happens for the solutions~\eqref{luninmathurtwocharge} dual to the highest weight state). The matching of the dual CFT and the gravitational descriptions of $|\psi_0\rangle$ was discussed in detail in~\cite{Kanitscheider:2007wq}.

\section{Solution generating technique}\label{section:generating}

\subsection{General solution}
\label{section:Warner}
A powerful way to solve the system of equations described above has been found in~\cite{Niehoff:2013kia}: one can express all the geometric data (apart from some of the components of $\omega$) algebraically in terms of ``generalized harmonic functions'' for the operator $\cald$, in much the same way as $v$-independent solutions are expressed in terms of ordinary harmonic functions on $\mathbb{R}^3$~\cite{Bena:2005va,Berglund:2005vb}. The results of~\cite{Niehoff:2013kia} apply to a restricted ansatz where the fields $Z_4, a_4, \delta_2$ have been set to zero: we will generalise here those results to our more general setting. 

Start from a hyperkahler 4D base of the Gibbons-Hawking form:
\be
d s^2_4 = V^{-1}(d\tau+A)^2 + V d s^2_3\,,
\ee
where $\tau$ is a particular direction in $\mathbb{R}^4$, $d s^2_3$ is the flat metric on $\mathbb{R}^3$, $V$ is a harmonic function on $\mathbb{R}^3$ and $A$ a 1-form on $\mathbb{R}^3$ related to
$V$ by
\be
*_3\! d_3 A = d_3 V\,,
\ee
with $d_3$ the differential on $\mathbb{R}^3$ and $*_3$ the Hodge dual associated with $d s^2_3$. In our applications $d s^2_4$ will be just the flat $\mathbb{R}^4$ metric, which corresponds to the choice $V=\frac{1}{\rho}$ (with $\rho$ the radial coordinate of $\mathbb{R}^3$), but the results of this subsection apply more generally for any GH potential $V$. 

A $v$-independent $\beta$ in this metric has the form
\be
\beta = \frac{K_3}{V} (d\tau+A) + \xi\,,
\ee
where $K_3$ is a harmonic function on $\mathbb{R}^3$ and the 1-form $\xi$ satisfies
\be
*_3 \!d_3\, \xi = - d_3 K_3\,.
\ee 

The system of equations for $Z_1$, $\Theta_2$ can be solved as
\begin{subequations}
\be\label{Theta2harm}
\Theta_2 = \cald \Bigl(\frac{K_2}{V}\Bigr) \wedge (d\tau+A) + *_4 \Bigl[\cald \Bigl(\frac{K_2}{V}\Bigr)
\wedge (d\tau+A) \Bigr]\,,
\ee
\be\label{Z1harm}
Z_1 = L_1 + \frac{K_2\,K_3}{V}\,.
\ee
\end{subequations}
$K_2$ and $L_1$ are ``generalized harmonic functions'' with respect to the differential $\cald$:
\be
*_4 \cald  *_4 \cald K_2 = *_4 \cald *_4 \cald L_1=0\,,
\ee
and have to satisfy
\be\label{K2L1}
\partial_\tau K_2 + \partial_v L_1=0\,.
\ee
Note that this is not the most general solution, but it is the one which will be relevant for most of our  applications. We will describe the general solution and one application in the Appendix.

Analogously one can solve for $Z_2$, $\Theta_1$:
\begin{subequations}
\be\label{Theta1harm}
\Theta_1 = \cald \Bigl(\frac{K_1}{V}\Bigr) \wedge (d\tau+A) + *_4 \Bigl[\cald \Bigl(\frac{K_1}{V}\Bigr)
\wedge (d\tau+A) \Bigr]\,,
\ee
\be\label{Z2harm}
Z_2 = L_2 + \frac{K_1\,K_3}{V}\,,
\ee
\end{subequations}
with
\be
*_4 \cald  *_4 \cald K_1 = *_4 \cald *_4 \cald L_2=0\,,
\ee
\be\label{K1L2}
\partial_\tau K_1 + \partial_v L_2=0\,,
\ee
and $Z_4$, $\Theta_4$:
\begin{subequations}
\be
\Theta_4 = \cald \Bigl(\frac{K_4}{V}\Bigr) \wedge (d\tau+A) + *_4 \Bigl[\cald \Bigl(\frac{K_4}{V}\Bigr)
\wedge (d\tau+A) \Bigr]\,,
\ee
\be\label{Z4harm}
Z_4 = L_4 + \frac{K_4\,K_3}{V}\,,
\ee
\end{subequations}
with
\be
*_4 \cald  *_4 \cald K_4 = *_4 \cald *_4 \cald L_4=0\,,
\ee
\be\label{K4L4}
\partial_\tau K_4 + \partial_v L_4=0\,.
\ee
The solution for $\mathcal{F}$ can be written as
\be\label{calFharm}
\mathcal{F}= L_3 + \frac{K_4^2 -K_1\,K_2 }{V}\,,
\ee
where $L_3$ is generalized harmonic
\be
*_4 \cald  *_4 \cald L_3=0\,,
\ee
and finally $\omega$ is given by
\be
\omega = \mu\,(d\tau+A) + \zeta\,,
\ee
where
\be\label{muharm}
\mu = M + \frac{L_1 \,K_1+L_2\,K_2 - L_3\,K_3 - 2\,L_4\,K_4}{2\,V}+\frac{(K_1\,K_2-K_4^2)\,K_3 }{V^2}\,,
\ee
with
\be
*_4 \cald  *_4 \cald M=0\,.
\ee
The 1-form along $\mathbb{R}^3$ $\zeta$ is not determined algebraically but by solving the following system of differential equations
\begin{subequations}
\label{zetaeqs}
\begin{align}
&*_3 \cald_3\, \zeta + (V \partial_\tau - K_3\, \partial_v) \zeta = V \cald_3 M - M \cald_3 V +\frac{1}{2}\,\Bigl[K_1\cald_3 L_1 - L_1\cald_3 K_1 \nonumber\\
&+K_2\cald_3 L_2 - L_2\cald_3 K_2 - (K_3\cald_3 L_3 - L_3\cald_3 K_3) + 2 \,(K_4\cald_3 L_4 - L_4\cald_3 K_4) \Bigr]\,,\\
&*_3 \cald_3 *_3 \zeta + V^2 \partial_\tau \mu + K_3^2\,\partial_v \Bigl[\frac{Z_1\,Z_2\,V}{\alpha\,K_3^2}-\frac{\mu\,V}{K_3}\Bigr]=0\,.
\end{align}
\end{subequations}
We have denoted by $\cald_3$ the component of $\cald$ along $\mathbb{R}^3$:
\be
 \cald_3\equiv d_3 -A\,\partial_\tau - \xi\,\partial_v\,.
\ee
In general Eqs. (\ref{zetaeqs}) form a coupled system of partial differential equations in three unknowns, the components of $\zeta$. It is sometimes more convenient to deal with uncoupled 
equations: for this purpose one can go back to the original supergravity constraints (\ref{eqcalFomega}), which can be written in the form
\be\label{eqomegabis}
\cald \omega + *_4 \cald \omega =\Omega_2\,,\quad *_4 \cald *_4 \omega = \Omega_0\,.
\ee
$\Omega_2$ and $\Omega_0$ are a 2-form and a 0-form that include all the $\omega$-independent terms of Eqs. (\ref{eqcalFomega}) and are completely known once $ds^2_4$, $\beta$, $Z_1$, $Z_2$, $Z_4$, $\Theta_1$, $\Theta_2$, $\Theta_4$ and $\mathcal{F}$ have been computed.  Eqs. (\ref{eqomegabis}) imply 
\be\label{laplomega}
\cald*_4\cald *_4 \omega +*_4 \cald*_4\cald \omega +*_4 \cald^2\omega = -\cald_j\cald_j\, \omega  = \cald\Omega_0+*_4 \cald \Omega_2\,.
\ee
The ``generalized Laplacian'' $\cald_j\cald_j$ acts diagonally on the Cartesian components of $\omega$, and hence Eq. (\ref{laplomega}) forms a set of four uncoupled partial differential equations 
of the second order for $\omega_i$. In concrete computations this provides often the most practical way to solve for $\omega$.

\subsection{Generating solutions via chiral algebra transformations}
\label{section:chiralalgebra}

The method described in the previous subsection allows in principle to construct a large sub-family of geometries carrying the same charges and supercharges of the D1-D5-P system, those for which there exists a coordinate system where the 4D base and $\beta$ do not depend on $v$. The geometries describing black hole microstates should be dual to well-defined CFT states. In this subsection we will provide a technique to generate solutions in the above sub-family whose CFT dual states can be easily identified. 

The method starts from a 2-charge (D1-D5) geometry in the class described in Section~\ref{2cfullg}; as explained in Section~\ref{2cdl}, in the decoupling limit the geometry is asymptotically isomorphic to AdS$_3\times S^3\times T^4$. One can then act on the near-horizon solution with a chiral algebra transformation in the right-moving sector of the CFT: on the gravity side these transformations act as diffeomorphisms that do not vanish at the boundary of AdS, and thus might transform the dual state into a physically inequivalent one. Transformations in the right-moving sector 
in general preserve only half of the supersymmetries of the original 2-charge state, and transform the RR ground states into excited RR states that carry D1, D5 and momentum charge and preserve four supercharges. To identify the corresponding geometries with black hole microstates one would need to glue back the near-horizon solution to the asymptotically flat region --- by which  we mean a geometry that is asymptotically $\mathbb{R}^{1,4}\times S^1\times T^4$. This is the technically challenging step of the construction: if one naively tries to add back the ``1'' to $Z_1$ and $Z_2$, so as to restore the correct asymptotics, one generically violates the supergravity equations. To solve this problem we exploit the results described in the previous subsection. We rewrite the near-horizon geometry obtained by the previously described sequence of coordinate transformations in the form of the general ansatz (\ref{ansatzsummary}) and extract the various geometric data. In general the 4D metric $d s^2_4$ and $\beta$ will be $v$-dependent. However, as we will explicitly show below, there are cases in which $d s^2_4$ and $\beta$ are $v$-independent, 
at least in appropriate coordinates, and we will restrict to this case. Then, as explained in section~\ref{section:Warner}, the near-horizon geometry can be encoded into generalized harmonic functions. To construct a solution with the required asymptotically flat behavior one should replace
\be\label{add1}
L_1 \to L_1 +1\,,\quad L_2 \to L_2 +1\,,
\ee 
keeping all other  generalized harmonic functions unchanged. According to Eq.~(\ref{muharm}), this replacement will also change the 1-form $\omega$, in such a way that the supergravity equations are preserved. The change in $\omega$, however, will generically introduce unphysical Dirac-Misner singularities. To generate a solution that is asymptotically flat, solves the supergravity constraints, and is regular, one has to perform the transformation (\ref{add1}) and at the same time correct the coefficients of the various ``harmonic'' functions so as to satisfy all the following regularity requirements: Generically the function $V$ has poles, where the Gibbons-Hawking fiber $\tau$ degenerates; the functions $Z_1$, $Z_2$, $Z_4$, $\mathcal{F}$, $\frac{K_3}{V}$ and $\mu$ must be regular at the positions of these poles and $\mu+\frac{K_3}{V}$ must vanish at the same positions (the last condition guarantees that
the Dirac string singularities of $\beta$ and $\omega$ can be canceled by a shift of the coordinate $y$). Moreover the functions $Z_1$, $Z_2$, $K_3$ and $\mu$ (but not $\mathcal{F}$) might have poles at other positions (corresponding to the location of the profile $g_i(v)$ in the original 2-charge geometry); these poles generate a possible singularity in the 10D metric proportional to $(d\tau+A)^2$ whose coefficient is
\be\label{regularitySigmazero}
-\frac{2\,\alpha}{\sqrt{Z_1 Z_2}}\,\frac{K_3}{V}\,\Bigl(\mu+\frac{\mathcal{F}}{2}\,\frac{K_3}{V}\Bigr)+\frac{\sqrt{Z_1 Z_2}}{V}=\frac{\alpha}{\sqrt{Z_1 Z_2}\,V}\,(L_1 L_2-L_4^2 -2 K_3 M)\,.
\ee
One should require the finiteness of this coefficient. We will show in concrete examples that these regularity constraints uniquely fix the coefficients of the various ``harmonic'' functions and hence lead to a unique regular and asymptotically flat solution.

The technique of generating solutions via chiral algebra transformations was applied to the construction of black hole microstates in \cite{Mathur:2003hj,Mathur:2011gz,Mathur:2012tj,Shigemori:2013lta}, but only at the perturbative level. In \cite{Lunin:2012gp} an exact supergravity solution obtained by acting with the torus symmetry $U(1)^4_L$ was constructed; since the $U(1)^4_L$ transformations break the isotropy along $T^4$, the solution of \cite{Lunin:2012gp} does not fit into the class of solutions considered in this paper. Here we focus on the $S^3$ rotations $SU(2)_R$, though in principle our method could be applied to $SL(2,\mathbb{R})$ transformations as well. In the next section we will detail the construction of two different solutions along the lines outlined above.

\section{Two different classes of geometries}\label{section:examples}

\subsection{Geometries with a \texorpdfstring{$v$}{}-independent base} \label{gvind}
The solution of this subsection represents the non-linear extension of the perturbative 3-charge solution found in \cite{Mathur:2003hj}; the non-linear solution was already presented in \cite{Giusto:2013rxa}; we will give here the details of its construction. 

We follow the solution generating technique explained in section~\ref{section:chiralalgebra}: we start from the 2-charge geometry given in Eqs. (\ref{skenderistwocharge}), take the decoupling limit by replacing $Z_1\to Z_1-1$, $Z_2\to Z_2-1$, go to the NS sector via the coordinate redefinition 
(\ref{spectralflow}), and act with the finite rotation generated by $R^2$, using (\ref{rotg}). To simplify the computation,  we restrict here to the particular value $\chi^{(2)}_R= \pi$, for which the transformation in (\ref{rotg}) reduces to
\be
\theta \to \frac{\pi}{2}-\theta\,,\quad \phi \to  -\psi\,,\quad \psi \to -\phi\,.
\ee
The computation for generic values of $\chi^{(2)}_R$ is a bit more involved, as it requires a generalization of the formalism of section~\ref{section:Warner}; we will provide the details of the computation for generic $\chi^{(2)}_R$ in the Appendix. To obtain a geometry dual to a state in the RR sector we finally perform the inverse of the spectral flow transformation (\ref{spectralflow}). The geometry that results from this sequence of transformations is described by the following geometric data:
\begin{subequations}
\label{nhvind}
\allowdisplaybreaks
\begin{align}
\label{3chargenhfirst}
d s^2_4 &=  (r^2+a^2 \cos^2\theta)\Bigl(\frac{d r^2}{r^2+a^2}+d\theta^2\Bigr)+(r^2+a^2)\sin^2\theta\,d\phi^2+r^2 \cos^2\theta\,d\psi^2\,,\\
\beta &=  \frac{R\,a^2}{\sqrt{2}\,(r^2+a^2 \cos^2\theta)}\,(\sin^2\theta\, d\phi - \cos^2\theta\,d\psi)\,,\\
Z_1 &=  \frac{R^2}{Q_5} \frac{a^2+\frac{b^2}{2}}{r^2+a^2 \cos^2\theta}+\frac{R^2\, a^2\, b^2}{2\,Q_5}\,\cos2\hat v\,\frac{\cos^2\theta}{(r^2+a^2 \cos^2\theta)(r^2+a^2)}\,,\\
Z_2 &=  \frac{Q_5}{r^2+a^2 \cos^2\theta}\,,\quad a_1=0\,,\quad \gamma_2 = -Q_5\,\frac{(r^2+a^2)\,\cos^2\theta}{r^2+a^2\cos^2\theta}\,d\phi\wedge d\psi\,,\\
Z_4 &=  R\, a\, b\,\cos\hat v \,\frac{\cos\theta}{\sqrt{r^2+a^2}\,(r^2+a^2 \cos^2\theta)}\,,\quad a_4 =0\,,\\
\delta_2 &= R\, a\, b\,\frac{r}{\sqrt{r^2+a^2}}\,\Bigl[\cos\hat v\,\sin\theta\,\Bigl(\frac{d r\wedge d\theta}{r^2+a^2}+\frac{r\,\sin\theta\cos\theta}{r^2+a^2 \cos^2\theta}\,d\phi\wedge 
d\psi\Bigr)\nonumber\\
&\qquad - \sin\hat v \Bigl(\frac{\cos\theta}{r^2+a^2}\,d r\wedge d\psi + \frac{\sin\theta}{r}\,d\theta\wedge d\phi\Bigr)\Bigr]\,,\label{delta2NH}\\
\omega &= \frac{R\,a^2}{\sqrt{2}\,(r^2+a^2 \cos^2\theta)}\,(\sin^2\theta\, d\phi + \cos^2\theta\, d\psi)\nonumber\\
&\qquad +\frac{R\,b^2}{\sqrt{2}}\frac{(r^2+a^2)\sin^2\theta\,d\phi+r^2\cos^2\theta\,d\psi}{(r^2+a^2)\,(r^2+a^2\cos^2\theta)}\,,\\\label{3chargenhlast}
\mathcal{F} &=  -\frac{b^2}{r^2+a^2}\,,
\end{align}
\end{subequations}
with
\be
\hat v = \frac{\sqrt{2}\,v}{R} -\psi\,. 
\ee
Note that the 4D metric $d s^2_4$ and $\beta$ have been left unchanged by the transformations, and in particular they are still $v$-independent. This allows us to apply the formalisms of section~\ref{section:Warner}: we will thus proceed to extract the generalized harmonic functions associated with the above solution. In our coordinates,\footnote{We could pass to coordinates, $(\rho,\eta,\varphi)$, where the 3-dimensional part of the Gibbons-Hawking metric $d s^2_3$ is explicitly flat $\mathbb{R}^3$:
\be
\rho = \frac{r^2+a^2\sin^2\theta}{4}\,,\quad \cos\eta =\frac{2r^2 \cos^2\theta}{r^2+a^2\sin^2\theta}-1\,,\quad \varphi=\psi-\phi\,, 
\ee
but we find it more convenient to continue using our original coordinates.} the Gibbons-Hawking fiber $\tau$, the potential $V$ for flat $\mathbb{R}^4$ and the associated 1-form $A$ are
\be\label{GHds4}
\tau=\psi+\phi\,,\quad V = \frac{4}{r^2+a^2\sin^2\theta}\,,\quad A = \Bigl(\frac{2r^2 \cos^2\theta}{r^2+a^2\sin^2\theta}-1\Bigr)(d\psi-d\phi)\,.
\ee
From $\beta$ we read off
\begin{subequations}\label{GHbeta}
\begin{align}
K_3 &= \sqrt{2}\,R\,\Bigl(\frac{1}{r^2+a^2\cos^2\theta}-\frac{1}{r^2+a^2\sin^2\theta}\Bigr)\,,\\
\xi& = \frac{R\,a^2}{\sqrt{2}}\frac{2\,r^2+a^2}{(r^2+a^2\sin^2\theta)(r^2+a^2\cos^2\theta)}\sin^2\theta\cos^2\theta\,(d\phi-d\psi)\,.
\end{align}
\end{subequations}
From $Z_2$, $a_1$, $\gamma_2$ one immediately sees that $\Theta_1=0$ and thus
\be
L_2 =  \frac{Q_5}{r^2+a^2 \cos^2\theta}\,,\quad K_1=0\,.
\ee
From the form of $Z_1$ one deduces that 
\be
L_1 =  \frac{R^2}{Q_5} \frac{a^2+\frac{b^2}{2}}{r^2+a^2 \cos^2\theta}+ \ell_1\cos2\hat v\,,
\ee
where $\ell_1$ is a function of only $r$ and $\theta$. The constraint (\ref{K2L1}) then implies
\be
K_2 =\frac{2\sqrt{2}}{R}\,\ell_1\cos2\hat v\,.
\ee
From (\ref{Z1harm}) one then deduces
\be
K_2  = \frac{\sqrt{2}\,a^2 \,b^2 R}{Q_5}\,\frac{\cos^2\theta}{(r^2+a^2)(r^2+a^2\sin^2\theta)}\,\cos 2\hat v\,,
\ee
\be
L_1 = \frac{R^2}{Q_5} \frac{a^2+\frac{b^2}{2}}{r^2+a^2 \cos^2\theta}+\frac{R^2\, a^2\, b^2}{2\,Q_5}\,\frac{\cos^2\theta}{(r^2+a^2)(r^2+a^2\sin^2\theta)}\,\cos 2\hat v\,.
\ee
The above value of $K_2$ and Eq. (\ref{Theta2harm}) allow us to compute 
\begin{equation}
\begin{aligned}
\Theta_2&= -\frac{\sqrt{2}\,R\,a^2\,b^2}{Q_5}\,\frac{r\,\cos\theta}{r^2+a^2}\Bigl[\sin2\hat v\,\sin\theta\,\Bigl(\frac{d r\wedge d\theta}{r^2+a^2}+\frac{r\,\sin\theta\cos\theta}{r^2+a^2 \cos^2\theta}\, d\phi\wedge d\psi\Bigr) \\
& \qquad +\cos2\hat v\Bigl(\frac{\cos\theta}{r^2+a^2}\,d r\wedge d\psi + \frac{\sin\theta}{r}\, d\theta\wedge d\phi\Bigr) \Bigr]\,.
\end{aligned}
\end{equation}
One can check that Eqs. (\ref{eqZ1Th2}) are indeed satisfied.

Similarly from $Z_4$ and the constraint (\ref{K4L4}) one finds
\be
K_4 = 2\sqrt{2}\,a\, b\,\frac{\cos\theta}{\sqrt{r^2+a^2}\,(r^2+a^2\sin^2\theta)}\,\cos\hat v\,,
\ee
\be
L_4 = R\,a\, b\, \frac{\cos\theta}{\sqrt{r^2+a^2}(r^2+a^2\sin^2\theta)}\,\cos\hat v\,.
\ee
One can verify that the $\Theta_4$ computed from the $K_4$ above equals $\dot{\delta}_2$, with $\delta_2$ given in (\ref{delta2NH}). 

The value of $L_3$ is extracted from $\mathcal{F}$:
\be
L_3=-\frac{b^2}{r^2+a^2\sin^2\theta}\,\Bigl[1+\frac{a^2\,\cos^2\theta}{r^2+a^2}\,\cos2\hat v\Bigr]\,.
\ee
Finally from $\omega$ one derives $M$:
\begin{align}
M=&\frac{R \,a^2}{2\sqrt{2}(r^2+a^2\cos^2\theta)}+\frac{R\,b^2}{4\sqrt{2}}\Bigl(\frac{1}{r^2+a^2\cos^2\theta}+\frac{1}{r^2+a^2\sin^2\theta}\Bigr)\nonumber\\
&+\frac{R\,a^2\,b^2}{4\sqrt{2}}\frac{\cos^2\theta}{(r^2+a^2)(r^2+a^2\sin^2\theta)}\,\cos2\hat v\,.
\end{align}
It is of course a check of the correctness of our calculation that $V$, $K_I$, $L_I$ and $M$ are all annihilated by the generalized laplacian $*_4\cald*_4\cald$.

We can now modify the generalized harmonic functions listed above in such a way that the modified geometry be asymptotically flat and regular; the modifications should be negligible in the near-horizon limit (\ref{eq:declim}). As explained in section~\ref{section:chiralalgebra}, the minimal modification $L_{1,2}\to 1+L_{1,2}$ does not work: this would also generate a variation of $\mu$:
\be\label{eq516}
\mu \to \mu + \frac{K_2}{2 V}\,,
\ee
and since $\frac{K_2}{2 V}$ does not vanish when $r^2+a^2\sin^2\theta \to 0$, this would spoil the regularity requirement that $\mu+\frac{K_3}{V}$ vanish when the Gibbons-Hawking fiber degenerates,
which happens when $r^2+a^2\sin^2\theta \to 0$.  Hence further modifications are necessary, and these are determined by the regularity conditions described in section~\ref{section:chiralalgebra}. In the present case the only potential singularities might come from the locus  $r^2+a^2\sin^2\theta \to 0$, where as we said the coordinate $\tau$ degenerates, or from the locus $r^2+a^2\cos^2\theta \to 0$, which coincides with the $\mathbb{R}^4$ projection of the profile $g_A(v)$. By inspection one can see that all the regularity constraints can be satisfied by multiplying the near-horizon value of $K_2$ by the factor $\frac{Q_5}{Q_5+a^2}$, which trivializes in the limit (\ref{eq:declim}), as required. The constraint (\ref{K2L1}) implies that also the $v$-dependent part of $L_1$ has to be multiplied by the same factor. So the generalized harmonic functions of the asymptotically flat solution are
\begin{subequations}
  \begin{align}
L_2 &=1 + \frac{Q_5}{r^2+a^2 \cos^2\theta}\,,
\\
K_2 & = \frac{\sqrt{2}\,a^2 \,b^2 R}{Q_5+a^2}\,\frac{\cos^2\theta}{(r^2+a^2)(r^2+a^2\sin^2\theta)}\,\cos 2\hat v\,,
\\
L_1 &= \frac{R^2}{Q_5} \frac{a^2+\frac{b^2}{2}}{r^2+a^2 \cos^2\theta}+\frac{R^2\, a^2\, b^2}{2\,(Q_5+a^2)}\,\frac{\cos^2\theta}{(r^2+a^2)(r^2+a^2\sin^2\theta)}\,\cos 2\hat v\,,
  \end{align}
\end{subequations}
with all other functions left invariant. The relations given in section~\ref{section:Warner} then allow to algebraically reconstruct all the geometric data apart from $\zeta$, the $\mathbb{R}^3$ part of $\omega$, for which one needs to solve the system of differential equations (\ref{zetaeqs}). We found it computationally easier to tackle the second order equations (\ref{laplomega}). The symmetries of the solution motivate the ansatz
\begin{align}
\omega&=\sin2\hat v\Bigl[\Bigl(h_1\cos\theta +h_2\frac{r\sin\theta}{\sqrt{r^2+a^2}}\Bigr)d r+\Bigl(-h_1 r\sin\theta +h_2 \sqrt{r^2+a^2}\cos\theta \Bigr)d\theta\Bigr]\\
&-\cos2\hat v \Bigl[h_2 \sqrt{r^2+a^2}\sin\theta \,d\phi +h_1 r \cos\theta \,d\psi\Bigr]+h_\phi \,d\phi+h_\psi\,d\psi\,,
\end{align}
where $h_1,h_2,h_\phi,h_\psi$ depend only on $r$ and $\theta$ and satisfy uncoupled second order partial differential equations, that can be solved. 

The final result is
\begin{subequations}
\label{3chargefin}
\allowdisplaybreaks
\begin{align}
d s^2_4 =& \,(r^2+a^2 \cos^2\theta)\Bigl(\frac{d r^2}{r^2+a^2}+d\theta^2\Bigr)+(r^2+a^2)\sin^2\theta\,d\phi^2+r^2 \cos^2\theta\,d\psi^2\,,\\
\beta=& \, \frac{R\,a^2}{\sqrt{2}\,(r^2+a^2 \cos^2\theta)}\,(\sin^2\theta\,d\phi - \cos^2\theta\,d\psi)\,,\\
Z_1=&\, 1+ \frac{R^2}{Q_5} \frac{a^2+\frac{b^2}{2}}{r^2+a^2 \cos^2\theta}+\frac{R^2\, a^2\, b^2}{2\,(Q_5+a^2)}\, \frac{\cos2\hat v\,\cos^2\theta}{(r^2+a^2 \cos^2\theta)(r^2+a^2)}\,,\\
Z_2=& \,1 + \frac{Q_5}{r^2+a^2 \cos^2\theta}\,,\quad a_1=0\,,\\
Z_4=&\, R\, a\, b\,\cos\hat v \,\frac{\cos\theta}{\sqrt{r^2+a^2}\,(r^2+a^2 \cos^2\theta)}\,,\quad a_4 =0\,,\\
\delta_2=& R\, a\, b\,\frac{r}{\sqrt{r^2+a^2}}\,\Bigl[\cos\hat v\,\sin\theta\,\Bigl(\frac{d r\wedge d\theta}{r^2+a^2}+\frac{r\,\sin\theta\cos\theta}{r^2+a^2 \cos^2\theta}\,d\phi\wedge 
d\psi\Bigr)\nonumber\\
& - \sin\hat v \Bigl(\frac{\cos\theta}{r^2+a^2}\,d r\wedge d\psi + \frac{\sin\theta}{r}\,d\theta\wedge d\phi\Bigr)\Bigr]\,,
\\
\omega = &  \frac{R\,a^2}{\sqrt{2}\,(r^2+a^2 \cos^2\theta)}\,(\sin^2\theta\,d\phi + \cos^2\theta\,d\psi)\nonumber\\
&+\frac{R\,b^2}{\sqrt{2}}\frac{(r^2+a^2)\sin^2\theta\,d\phi+r^2\cos^2\theta\,d\psi}{(r^2+a^2)\,(r^2+a^2\cos^2\theta)}\nonumber\\
&-\frac{R\,a^2\,b^2}{2\sqrt{2}\,(Q_5+a^2)}\,\Bigl[\cos2\hat v\,\frac{a^2 \sin^2\theta\, d\phi-r^2 \,d\psi}{(r^2+a^2)\,(r^2+a^2\cos^2\theta)}\,\cos^2\theta\nonumber\\
&+\sin2\hat v\,\frac{r \cos\theta\,d r - (r^2+a^2)\sin\theta\, d\theta}{(r^2+a^2)^2}\,\cos\theta\Bigr] \,,\\
\mathcal{F} = & -\frac{b^2}{r^2+a^2}\,.
\end{align}
\end{subequations}
By construction the geometry defined above is asymptotically flat, completely regular, carries the same charges and supercharges as the D1-D5-P black hole, and reduces in the near-horizon region to the microstate obtained by acting with the R-symmetry rotation (\ref{rotg}) on the RR ground state (\ref{eq:Esk}). In section~\ref{svind} we compute the average values of the R-charges $J^3_0$, $\tilde J^3_0$ and of the momentum operator $L_0-\tilde L_0$ on this microstate in the orbifold CFT, and compare them with the asymptotic charges derived from the geometry (\ref{3chargefin}) in section~\ref{section:CFTmatching}.

\subsection{Geometries with a \texorpdfstring{$v$}{}-dependent base}  \label{gvd}
In this subsection we construct a new 3-charge microstate that differs from the previous one in two respects: the starting 2-charge solution is different and the transformation one applies in the NSNS sector does not belong to the R-charge group $SU(2)$ but to its affine extension. It represents the non-linear completion of the solution discussed in section 4 of \cite{Mathur:2012tj}. From a technical point of view, the example of this subsection is complicated by the fact that the $d s^2_4$ and $\beta$ one obtains after the chiral algebra transformation are $v$-dependent; one can still apply the formalism of section~\ref{section:Warner} at the price of working in a system of coordinates where the metric in the asymptotic region doe not explicitly reduce to the flat space
$M^{1,4}\times S^1\times T^4$.\footnote{The trick of working in a non-asymptotically flat coordinate frame to simplify the solution of the equations of motion has been employed several times in the past \cite{Garfinkle:1990jq,Callan:1995hn,Dabholkar:1995nc,Lunin:2012gp}.}
 
The seed 2-charge geometry is the Lunin-Mathur geometry with circular profile given in (\ref{luninmathurtwocharge}). As usual the near-horizon limit is obtained by replacing $Z_1$ and $Z_2$ with
\be
Z_1^\mathrm{nh}=\frac{Q_1}{r^2+a^2\cos^2\theta}\,,\quad Z_2^\mathrm{nh}=\frac{Q_5}{r^2+a^2\cos^2\theta}\,.
\ee
In this limit, and after going to the NSNS sector via (\ref{spectralflow}), the geometry reduces to
AdS$_3\times S^3\times T^4$. On this geometry we want to act with a transformation corresponding to the affine $SU(2)$ generator $J^3_{-n}$ (where we are taking $n>0$): this was identified in \cite{Mathur:2012tj} with the diffeomorphism\footnote{With respect to the conventions of \cite{Mathur:2012tj}, we have $\phi\to -\phi$.}
\be
\phi\to \phi + \frac{1}{2}\,\hat\epsilon\,e^{-i\,\frac{n\sqrt{2}\,v}{R}}\,,\quad \psi\to \psi + \frac{1}{2}\,\hat\epsilon\,e^{-i\,\frac{n\sqrt{2}\,v}{R}}\,,
\ee
where the factor $1/2$ descends from the $1/2$ factor in (\ref{tildeJ3}).
As we are working at non-linear order in $\hat\epsilon$, to generate a real geometry we should act with a real version of the above transformation; we choose
\be\label{J3minusn}
\phi\to \phi - \hat\epsilon\,\sin\Bigl(\frac{n\sqrt{2}\,v}{R}\Bigr)\,,\quad \psi\to \psi - \hat\epsilon\,\sin\Bigl(\frac{n\sqrt{2}\,v}{R}\Bigr)\,.
\ee
The near-horizon geometry for the microstate in the RR sector is obtained after the inverse of the spectral flow (\ref{spectralflow}). Since spectral flow commutes with the transformation (\ref{J3minusn}),
the final geometry is equivalent to the one obtained by acting with (\ref{J3minusn}) directly on the 2-charge geometry in the RR sector. We can formally rewrite (\ref{J3minusn}) as
\be\label{coordinateshift}
x^i \to x^i - f^i(v)\,,
\ee 
where the only non-trivial components of the ``profile'' $f^i(v)$ are
\be\label{profileJ3}
f^\phi(v)=f^\psi(v) =  \hat\epsilon\,\sin\Bigl(\frac{n\sqrt{2}\,v}{R}\Bigr)\,,
\ee 
while in the Gibbons-Hawking coordinates introduced in the previous section the only non-trivial component is 
\begin{equation}
  \label{eq:ftau}
  f^\tau(v) = 2\, \hat \epsilon \sin\Bigl(\frac{n\sqrt{2} v}{R}\Bigr)~.
\end{equation}
The action of a transformation of the form (\ref{coordinateshift}) on a general solution in the ansatz (\ref{ansatzsummary}) was already worked out in Appendix B of  \cite{Giusto:2012jx}. We can apply those transformation rules to the near-horizon limit of the solution (\ref{luninmathurtwocharge}) and obtain a geometry described by the following geometric data:
\begin{subequations}
\label{nearhorizonJ3}
\begin{align}
d \hat s^2_4 &=(1-\beta_k\,\dot{f}^k)\,d x^i d x^i + (\beta_j\,\dot{f}_i +\beta_i\,\dot{f}_j )\,d x^i d x^j +\frac{\beta_i\,\beta_j}{1-\beta_k\,\dot{f}^k}\,|\dot{f}|^2\, d x^i\,d x^j  \,,\label{4Dmetricprofile}\\
\hat \beta&=\frac{\beta}{1-\beta_k \, \dot{f}^k}\,,\label{betaprofile} \\
\hat Z_1&= \frac{Z^\mathrm{nh}_1}{1-\beta_k \, \dot{f}^k}\,,\quad \hat Z_2= \frac{Z^\mathrm{nh}_2}{1-\beta_k \, \dot{f}^k}\,,\label{Z1Z2profile}\\
\hat \omega&=\omega+\beta\,\Bigl(\frac{\omega_l\,\dot{f}^l}{1-\beta_k \, \dot{f}^k} +\frac{Z^\mathrm{nh}_1 Z^\mathrm{nh}_2}{(1-\beta_k \, \dot{f}^k)^2}\,|\dot{f}^2|\Bigr) + \frac{Z^\mathrm{nh}_1 Z^\mathrm{nh}_2}{1-\beta_k \, \dot{f}^k}\,\dot{f}_i\,d x^i\,,\label{omegaprofile}\\
\widehat{\mathcal{F}}&=-2\,\omega_k\,\dot{f}^k-\frac{Z^\mathrm{nh}_1 Z^\mathrm{nh}_2}{1- \beta_k \, \dot{f}^k}\,|\dot{f}|^2\,,\label{calFprofile}\\
\hat a_1&= Z^\mathrm{nh}_2\,\Bigl(\dot{f}_i\,d x^i+\frac{\beta}{1-\beta_k \, \dot{f}^k}\,|\dot{f}|^2\Bigr) -  \gamma_{2\,ij}\,d x^i\,\dot{f}^j\,,\label{a1profile}\\
\hat\gamma_2& = \gamma_2 + \gamma_{2\,ij}\,\dot{f}^i\,\frac{\beta}{1-\beta_k \, \dot{f}^k}\wedge d x^j\,.
\end{align}
\end{subequations}
The solution defined by the hatted quantities describes the near-horizon limit of a new 3-charge microstate. We now extend this 3-charge geometry to the asymptotically flat region.

Eqs. (\ref{4Dmetricprofile},\ref{betaprofile}) present a problem: due to $\dot{f}^i$, $d \hat s^2_4$ and $\hat \beta$ depend on $v$, and thus we cannot straightforwardly apply the formalism of section~\ref{section:Warner}. So if one tries to construct a geometry with flat asymptotics by the usual trick of adding a ``1'' to $\hat Z_1$ and $\hat Z_2$, one can see from the general supergravity equations given in \cite{Giusto:2013rxa} (cf. Eqs. (E.54a), (E.56a) of that reference) that  
one violates the supergravity constraints by terms proportional to $\dot{\hat\beta}$. There is an easy fix to this problem: one can simply add a ``1'' to $Z^\mathrm{nh}_1$ and $Z^\mathrm{nh}_2$ in Eq. (\ref{Z1Z2profile}) and obtain warp factors $\hat Z_1$ and $\hat Z_2$ that both have the right asymptotic limit and satisfy the supergravity constraints. This operation, however, generates further difficulties: Eqs. (\ref{omegaprofile},\ref{calFprofile},\ref{a1profile}) show that after inserting back the ``1'' in the warp factors $Z^\mathrm{nh}_1$ and $Z^\mathrm{nh}_2$, $\hat \omega$, $\widehat{\mathcal{F}}$ and $\hat a_1$ have the asymptotic limits
\be
\hat \omega \to \dot{f}_i\,d x^i\,,\quad \widehat{\mathcal{F}}\to - |\dot{f}|^2\,,\quad \hat a_1\to \dot{f}_i\,d x^i\,,
\ee
which are not the appropriate ones for an asymptotically flat geometry. One can restore the correct asymptotics by modifying $\omega$ and by introducing non-trivial values for the fields $\mathcal{F}$ and $a_1$ before the coordinate shift (\ref{coordinateshift}); preserving the supergravity constraints will also force a modification of $Z_1$ and $Z_2$. If we denote with a tilde these modified geometric data, the asymptotically flat extension of the 3-charge geometry (\ref{nearhorizonJ3}) has the form
\begin{subequations}
\label{shiftaction}
\begin{align}
d \hat s^2_4 &=(1-\beta_k\,\dot{f}^k)\,d x^i d x^i + (\beta_j\,\dot{f}_i +\beta_i\,\dot{f}_j )\,d x^i d x^j +\frac{\beta_i\,\beta_j}{1-\beta_k\,\dot{f}^k}\,|\dot{f}|^2\, d x^i\,d x^j  \,,\\
\hat \beta&=\frac{\beta}{1-\beta_k \, \dot{f}^k}\,, \\
\hat Z_1&= \frac{\tilde Z_1}{1-\beta_k \, \dot{f}^k}\,,\quad \hat Z_2= \frac{\tilde Z_2}{1-\beta_k \, \dot{f}^k}\,,\\
\hat \omega&=\tilde \omega+\beta\,\Bigl(\frac{\tilde\omega_l\,\dot{f}^l}{1-\beta_k \, \dot{f}^k} +\frac{\tilde Z_1 \tilde Z_2}{(1-\beta_k \, \dot{f}^k)^2}\,|\dot{f}^2|\Bigr) + \frac{\tilde Z_1 \tilde Z_2}{1-\beta_k \, \dot{f}^k}\,\dot{f}_i\,d x^i\,,\label{hatomega}\\
\widehat{\mathcal{F}}&=\widetilde{\mathcal{F}}\, (1-\beta_k \, \dot{f}^k)-2\,\tilde \omega_k\,\dot{f}^k-\frac{\tilde Z_1 \tilde Z_2}{1-\tilde \beta_k \, \dot{f}^k}\,|\dot{f}|^2\,,\label{hatcalF}\\
\hat a_1&=\tilde a_1\,(1-\beta_k \, \dot{f}^k) +\beta\,\tilde a_{1\, k}\,\dot{f}^k+\tilde Z_2\,\Bigl(\dot{f}_i\,d x^i+\frac{\beta}{1-\beta_k \, \dot{f}^k}\,|\dot{f}|^2\Bigr) - \tilde \gamma_{2\,ij}\,d x^i\,\dot{f}^j\,,\\
\hat\gamma_2& = \tilde\gamma_2 + \tilde\gamma_{2\,ij}\,\dot{f}^i\,\frac{\beta}{1-\beta_k \, \dot{f}^k}\wedge d x^j\,,
\end{align}
\end{subequations}
where one requires the following behavior for the ``tilded'' quantities  at large distances
\be\label{tildedasymptotics}
\tilde Z_1\to 1\,,\quad \tilde Z_2\to 1\,,\quad 
\tilde \omega \to -\dot{f}_i\,d x^i\,,\quad \widetilde{\mathcal{F}}\to - |\dot{f}|^2\,,\quad \tilde a_1\to -\dot{f}_i\,d x^i\,.
\ee
One can think of the solution associated with the tilded quantities, together with $\beta$ and the flat 
$d s^2_4$, as the geometry representing the 3-charge microstate in a system of coordinates where the asymptotically flat structure is not manifest. The change of coordinates (\ref{coordinateshift}), which transforms the tilded quantities into the hatted ones, brings the solution into an explicitly flat frame at asymptotic infinity. This procedure is analogous to the solution generating technique of \cite{Garfinkle:1990jq}, which was used in \cite{Callan:1995hn,Dabholkar:1995nc} to construct F1-P solutions starting from the static F1 solution: in an analogous way, we start from a D1-D5 solution and generate a D1-D5-P solution. There are however some differences in the two cases. In the solution of \cite{Callan:1995hn,Dabholkar:1995nc}, $\tilde \omega$, $\widetilde{\mathcal{F}}$ and $\tilde a_1$ could be taken equal to their asymptotic values specified in (\ref{tildedasymptotics}). In our present case such an ansatz would not solve the supergravity equations, due to the presence of a nontrivial $\beta$. The non-triviality of $\beta$, which originates from the KK-monopole dipole charge generated from the binding of D1 and D5 charges, thus represents the main technical obstacle in the construction of 3-charge microstates. To construct 
$\tilde Z_1$, $\tilde Z_2$, $\tilde \omega$, $\widetilde{\mathcal{F}}$ and $\tilde a_1$ we have to solve a non-trivial system of differential equations. Since in the ``tilded frame'' the 4D metric  $d s^2_4$ is flat and $\beta$ is $v$-independent, we can take advantage however of the framework of section~\ref{section:Warner}. 

Let us then express $d s^2_4$ and $\beta$ in Gibbons-Hawking form as in (\ref{GHds4}), (\ref{GHbeta}) and let us look for $\widetilde{\Theta}_1$ and $\widetilde{\Theta}_2$ of the form  (\ref{Theta2harm},\ref{Theta1harm}). By looking at the $v$-dependence of the profile in (\ref{profileJ3}), it is natural to guess that the corresponding ``generalized harmonic'' functions, $\tilde K_1$ and $\tilde K_2$, have to be proportional to $\cos\Bigl(\frac{n\sqrt{2}\,v}{R}\Bigr)$. Looking for solutions of the equation  $\cald *_4 \cald \tilde K_{1,2}=0$ with such a $v$-dependence one finds two possible solutions
\be\label{generalizedharmonicn}
\cos \Bigl(\frac{n\sqrt{2}\,v}{R}\Bigr)\,\Bigl(\frac{r}{\sqrt{r^2+a^2}}\Bigr)^n\,,\quad \cos \Bigl(\frac{n\sqrt{2}\,v}{R}\Bigr)\,\Bigl(\frac{\sqrt{r^2+a^2}}{r}\Bigr)^n\,.
\ee
The second solution is singular at $r=0$ and should be discarded. The overall coefficient is determined by the asymptotic boundary condition for $\tilde a_1$ in (\ref{tildedasymptotics}): on one side, we have
\be\label{dotf}
\dot{f}_i\, dx^i = \frac{n\,\sqrt{2}\,\hat \epsilon}{R}\,\cos\Bigl(\frac{n\sqrt{2}\,v}{R}\Bigr)\,(r^2\cos^2\theta\,d\psi+(r^2+a^2)\sin^2\theta\,d\phi)\,,
\ee
and on the other side the asymptotic value for the $\tau$-component of $\tilde a_{1}$ is
\be\label{tildea1approx}
(\tilde a_{1})_\tau \approx \frac{\tilde K_2}{V}\approx \frac{r^2}{4}\,\tilde K_2\,,
\ee
where the approximation above is valid for large $r$. Comparing (\ref{dotf}) and (\ref{tildea1approx}) with (\ref{tildedasymptotics}), one finds
\be
\tilde K_2 =  -\frac{2\,n\,\sqrt{2}\,\hat\epsilon}{R}\, \cos \Bigl(\frac{n\sqrt{2}\,v}{R}\Bigr)\,\Bigl(\frac{r}{\sqrt{r^2+a^2}}\Bigr)^n\,.
\ee
The symmetry of the equations under exchange of the indices 1 and 2 implies
\be
\tilde K_1 = \tilde K_2 \equiv \tilde K .
\ee

We can assume that the harmonic functions $L_1$ and $L_2$ are the same as in the original 2-charge geometry:
\be
L_1 = 1+ \frac{Q_1}{r^2+a^2 \cos^2\theta}\,,\quad L_2 = 1+ \frac{Q_5}{r^2+a^2 \cos^2\theta}\,.
\ee
Note that this is consistent with the constraints (\ref{K2L1},\ref{K1L2}) since $\tilde K_1$ and 
$\tilde K_2$ are $\tau$-independent. The relations (\ref{Z1harm}) and (\ref{Z2harm}) then allow 
us to find $\tilde Z_1$ and $\tilde Z_2$:
\be
\tilde Z_1 = L_1 + \frac{\tilde K\,K_3}{V}\,,\quad \tilde Z_2 = L_2 + \frac{\tilde K\,K_3}{V}\,.
\ee

It follows from the general expressions (\ref{calFharm}) and (\ref{muharm}) that
\begin{subequations}
\begin{align}
\widetilde{\mathcal{F}} &= \tilde L_3 -\frac{\tilde K^2}{V}\,,\label{tildecalF}\\
\tilde\mu &= \tilde M + \frac{(L_1+L_2) \,\tilde K - \tilde L_3\,K_3}{2\,V}+\frac{\tilde K^2\,K_3 }{V^2}\,,
\end{align}
\end{subequations}
where $\tilde L_3$ and $\tilde M$ are generalized harmonic functions that are determined by the requirements of asymptotic flatness and regularity.  Note that the leading order terms in the large distance expansion of $\widetilde{\mathcal{F}}$ and $\tilde \mu$ come, respectively, from $-\frac{\tilde K^2}{V}$ and $\frac{(L_1 +L_2)\,\tilde K }{2\,V}$ and are consistent with the required asymptotic limits (\ref{tildedasymptotics}). For the geometry to be asymptotically flat, also the subleading terms of $\widehat{\mathcal{F}}$ and $\hat\omega$ must vanish: by using the form  of the profile~\eqref{eq:ftau} and Eqs.~(\ref{hatomega}) and~(\ref{hatcalF}), we see that asymptotic flatness of the metric requires the following large distance limits for $\tilde L_3$ and $\tilde M$
\begin{subequations}
\begin{align}
\tilde L_3 &\to -\frac{2\,n^2\,\hat\epsilon^2}{R^2}\,(Q_1+Q_5+n\,a^2)\,\cos^2\Bigl(\frac{n\sqrt{2}\,v}{R}\Bigr)\,,\label{tildeL3as}\\
\tilde M &\to -\frac{n\,\hat\epsilon}{2\sqrt{2}\,R}\,(Q_1+Q_5+n\,a^2)\,\cos\Bigl(\frac{n\sqrt{2}\,v}{R}\Bigr)\,.\label{tildeMas}
\end{align}
\end{subequations}
Hence $\tilde L_3$ contains generalized harmonic functions proportional to $\cos\Bigl(\frac{2 n\sqrt{2}\,v}{R}\Bigr)$; as seen in (\ref{generalizedharmonicn}), there are two possibilities, proportional to
\be
\cos \Bigl(\frac{2n\sqrt{2}\,v}{R}\Bigr)\,\Bigl(\frac{r^2}{r^2+a^2}\Bigr)^n\,,\quad \cos \Bigl(\frac{2n\sqrt{2}\,v}{R}\Bigr)\,\Bigl(\frac{r^2+a^2}{r^2}\Bigr)^n\,.
\ee
Regularity of $\widetilde{\mathcal{F}}$ at $r=0$, together with the asymptotic limit (\ref{tildeL3as}), univocally implies
\be
\tilde L_3 =  -\frac{n^2\,\hat\epsilon^2}{R^2}\,(Q_1+Q_5+n\,a^2)\,\Bigl[1+\cos \Bigl(\frac{2n\sqrt{2}\,v}{R}\Bigr)\,\Bigl(\frac{r^2}{r^2+a^2}\Bigr)^n\Bigr]\,.
\ee
The regularity conditions for $\tilde\mu$ are
\be\label{mutildetheta0}
\tilde\mu+\frac{K_3}{V}\to 0\quad \mathrm{for}\quad r\to0\,,\,\theta\to 0
\ee 
and
\be\label{mutildethetapiby2}
-\frac{2}{\sqrt{\tilde Z_1 \tilde Z_2}}\,\frac{K_3}{V}\,\Bigl(\tilde\mu+\frac{\widetilde{\mathcal{F}}}{2}\,\frac{K_3}{V}\Bigr)+\frac{\sqrt{\tilde Z_1 \tilde Z_2}}{V}=\frac{L_1 L_2 -2 K_3 \tilde M }{\sqrt{\tilde Z_1 \tilde Z_2}\,V}\to \mathrm{finite}\,\,\, \mathrm{for}\,\,\, r\to0\,,\,\theta\to \frac{\pi}{2}\,.
\ee
The unique function $\tilde M$ that is a linear combination of generalized harmonic functions, has the asymptotic limit (\ref{tildeMas}) and  satisfies the regularity condition (\ref{mutildethetapiby2}) is
\be
\tilde M = \frac{Q_1 Q_5}{2\sqrt{2}\,R\,(r^2+a^2 \cos^2\theta)} -\frac{n\,\hat\epsilon}{2\sqrt{2}\,R}\,(Q_1+Q_5+n\,a^2)\,\cos\Bigl(\frac{n\sqrt{2}\,v}{R}\Bigr)\,\,\Bigl(\frac{r}{\sqrt{r^2+a^2}}\Bigr)^n\,.
\ee
The other regularity constraint (\ref{mutildetheta0}) requires that
\be
R^2 = \frac{Q_1 Q_5}{a^2}-\frac{1}{2}\,\hat\epsilon^2\,n^2 (Q_1 +Q_5+n\,a^2)\,,
\ee
which is a deformation of the radius relation (\ref{Q1Q5Ra}). The constraint above should be interpreted as the relation that determines the parameter $a$ in terms of the asymptotic physical quantities $R$, $Q_1$, $Q_5$ and the parameters of the perturbation $\hat\epsilon$, $n$:
\be
a^2 =\frac{\sqrt{(2R^2+\hat\epsilon^2\, n^2\, (Q_1+Q_5))^2 +8\, \hat\epsilon^2 \,n^3\, Q_1 Q_5}-(2R^2+\hat\epsilon^2\, n^2\, (Q_1+Q_5))}{8\, \hat\epsilon^2\, n^3}\,.
\ee
Note that the solution for $a^2$ is real and positve for any value of $\hat\epsilon$ and $n>0$: thus a regular and asymptotically flat solution exists in all the range of CFT parameters $\hat\epsilon$ and $n$. For small values of the perturbation one of course recovers the 2-charge relation:
\be
a^2 \to \frac{Q_1 Q_5}{R^2} \quad \mathrm{for}\quad \hat\epsilon\to 0\,.
\ee
For large values of the perturbation, $a^2$ decreases (with the asymptotic quantities $R$, $Q_1$ and $Q_5$ held finite):
\be
a^2 \to \frac{2\,Q_1 Q_5}{\hat\epsilon^2\,n^2\,(Q_1+Q_5)}\quad \mathrm{for}\quad \hat\epsilon\to \infty\,.
\ee
Since the radius of the $y$ circle in the ``throat'' of the geometry is inversely proportional to $a$:
\be
R_\mathrm{throat}= \frac{\sqrt{Q_1 Q_5}}{a}\,.
\ee
This agrees with the intuitive expectation that the momentum carrying perturbation is localised in the throat and that it is responsible for the expansions of the $S^1$ in that region.

The final task is to compute the $\mathbb{R}^3$ part of $\tilde\omega$.  We make the ansatz
\begin{align}
\tilde\omega &= h_\phi\,d\phi + h_\psi\,d\psi + \cos\Bigl(\frac{n\sqrt{2}\,v}{R}\Bigr) \Bigl[(g_1 -g_2) \sqrt{r^2+a^2}\sin\theta \,d\phi +(h_1-h_2) r \cos\theta \,d\psi\Bigr] \nonumber \\
&+ \sin\Bigl(\frac{n\sqrt{2}\,v}{R}\Bigr) \Bigl[\Bigl((h_1+h_2)\cos\theta +(g_1+g_2)\frac{r\sin\theta}{\sqrt{r^2+a^2}}\Bigr)d r \nonumber \\
&+\Bigl((g_1+g_2)\sqrt{r^2+a^2}\cos\theta -(h_1+h_2)r\sin\theta \Bigr)d \theta  \Bigr] \nonumber  \\
&+ \cos\Bigl(\frac{2 n\sqrt{2}\,v}{R}\Bigr) \Bigl[(g'_1 -g'_2) \sqrt{r^2+a^2}\sin\theta \,d\phi +(h'_1-h'_2) r \cos\theta \,d\psi\Bigr] \nonumber  \\
&+ \sin\Bigl(\frac{2n\sqrt{2}\,v}{R}\Bigr) \Bigl[\Bigl((h'_1+h'_2)\cos\theta +(g'_1+g'_2)\frac{r\sin\theta}{\sqrt{r^2+a^2}}\Bigr)d r \nonumber \\
&+\Bigl((g'_1+g'_2)\sqrt{r^2+a^2}\cos\theta -(h'_1+h'_2)r\sin\theta \Bigr)d \theta  \Bigr]\,,
\end{align}
with $h_\phi$, $h_\psi$, $h_i$, $g_i$, $h'_i$, $g'_i$ functions of $r$ and $\theta$; substituting this ansatz in (\ref{laplomega}) one obtains uncoupled second order partial differential equations for these functions, that can be solved. 

The complete solution is specified by the following geometric data
\begin{subequations}
\label{solutiondatafinal}
\allowdisplaybreaks
 \begin{align}
d s^2_4 &= (r^2+a^2 \cos^2\theta)\Bigl(\frac{d r^2}{r^2+a^2}+d \theta^2\Bigr)+(r^2+a^2)\sin^2\theta\, d \phi^2+r^2 \cos^2\theta\,d \psi^2\,,\\
\beta&= \frac{R\,a^2}{\sqrt{2}\,(r^2+a^2\cos^2\theta)}\,(\sin^2\theta\,d\phi - \cos^2\theta\,d\psi)\,,\\
\tilde Z_1&=1+\frac{Q_1}{r^2+a^2\cos^2\theta}+ n\, \hat\epsilon \cos \Bigl(\frac{n\sqrt{2}\,v}{R}\Bigr)\,\Bigl(\frac{r}{\sqrt{r^2+a^2}}\Bigr)^n\,\frac{a^2 \cos2\theta}{r^2+a^2\cos^2\theta}\,,\\
\tilde Z_2&=1+\frac{Q_5}{r^2+a^2\cos^2\theta}+ n\, \hat\epsilon \cos \Bigl(\frac{n\sqrt{2}\,v}{R}\Bigr)\,\Bigl(\frac{r}{\sqrt{r^2+a^2}}\Bigr)^n\,\frac{a^2 \cos2\theta}{r^2+a^2\cos^2\theta}\,,\\\label{Thetatilde}
\tilde\Theta&=-\frac{n \sqrt{2} \hat\epsilon}{R\,r}\Bigl(\frac{r}{\sqrt{r^2+a^2}}\Bigr)^n\Bigl[
\cos \Bigl(\frac{n \sqrt{2}\,v}{R}\Bigr) (2 r^2+n a^2)\sin\theta (\sin\theta\,dr\wedge d\phi-r\cos\theta\,d\theta\wedge d\psi)\nonumber\\
&+\cos \Bigl(\frac{n \sqrt{2}\,v}{R}\Bigr)\,r^2 (2 r^2+(n+2) a^2) \cos\theta\,\Bigl(\frac{\cos\theta}{r^2+a^2}\,dr\wedge d\psi + \frac{\sin\theta}{r}\,d\theta\wedge d\phi\Bigr)\nonumber\\
&+\sin \Bigl(\frac{n \sqrt{2}\,v}{R}\Bigr) n\,a^2 (2 r^2+a^2) \sin\theta\cos\theta\,\Bigl(\frac{dr\wedge d\theta}{r^2+a^2}+\frac{r\sin\theta\cos\theta}{r^2+a^2\cos^2\theta}\,d\phi\wedge d\psi\Bigr)\Bigr]\,,\\
\tilde\omega&=\frac{R\,a^2}{\sqrt{2}\,(r^2+a^2\cos^2\theta)}\,(\sin^2\theta\,d\phi + \cos^2\theta\,d\psi)\nonumber\\
& -\frac{n\,\hat\epsilon}{\sqrt{2}\,R}\,\Bigl(\frac{r}{\sqrt{r^2+a^2}}\Bigr)^n \Bigl[\cos \Bigl(\frac{\sqrt{2}\,n\,v}{R}\Bigr)\Bigl(2 a^2 \sin^2\theta d\phi\nonumber\\
&+\Bigl((Q_1+Q_5)\frac{2r^2+a^2}{r^2+a^2\cos^2\theta}+2r^2+n a^2\Bigr)
(\sin^2\theta d\phi+\cos^2\theta d\psi)\Bigr)\nonumber\\
&+a^2\sin \Bigl(\frac{\sqrt{2}\,n\,v}{R}\Bigr)\,\Bigl(\frac{Q_1+Q_5+n\, r^2\cos 2\theta +n\, a^2 \cos^2\theta}{r(r^2+a^2)} dr-n\, \sin2\theta\,d\theta\Bigr)\Bigr]\nonumber\\
&+\frac{n^2 a^2 \hat\epsilon^2 (Q_1+Q_5+n\,a^2)}{\sqrt{2}\,R}\,\frac{\sin^2\theta}{r^2+a^2\cos^2\theta}\,d\phi \nonumber\\
 & -\frac{n^2 a^2 \hat\epsilon^2}{\sqrt{2}\,R}\,\Bigl(\frac{r^2}{r^2+a^2}\Bigr)^n\,\cos2\theta\,\frac{(r^2+a^2)\sin^2\theta\, d\phi+ r^2\cos^2\theta \,d\psi}{r^2+a^2\cos^2\theta}\nonumber\\
&+\frac{n^2 a^2 \hat\epsilon^2 (Q_1+Q_5+n\,a^2)}{2\sqrt{2}\,R}\Bigl(\frac{r^2}{r^2+a^2}\Bigr)^n\Bigl[\cos \Bigl(\frac{2\sqrt{2}\,n\,v}{R}\Bigr)\,\frac{\sin^2\theta\,d\phi-\cos^2\theta\,d\psi}{r^2+a^2\cos^2\theta}\nonumber\\
&\qquad-\sin \Bigl(\frac{2\sqrt{2}\,n\,v}{R}\Bigr)\,\frac{dr}{r(r^2+a^2)}\Bigr]\nonumber\\
&-\frac{n^2 a^2 \hat\epsilon^2}{\sqrt{2}\,R}\,\cos \Bigl(\frac{2\sqrt{2}\,n\,v}{R}\Bigr) \Bigl(\frac{r^2}{r^2+a^2}\Bigr)^n\cos2\theta\,\frac{(r^2+a^2)\sin^2\theta\,d\phi+r^2\cos^2\theta\,d\psi}{r^2+a^2\cos^2\theta} \,,\\
\widetilde{\mathcal{F}}&= -\frac{n^2\,\hat\epsilon^2}{R^2}\,(Q_1+Q_5+n\,a^2)\,\Bigl[1+\cos \Bigl(\frac{2n\sqrt{2}\,v}{R}\Bigr)\,\Bigl(\frac{r^2}{r^2+a^2}\Bigr)^n\Bigr]\\
& -
\frac{2\,n^2\,\hat\epsilon^2}{R^2}\, \cos^2 \Bigl(\frac{n\sqrt{2}\,v}{R}\Bigr)\,(r^2+a^2\sin^2\theta)\,\Bigl(\frac{r^2}{r^2+a^2}\Bigr)^n\,.
 \end{align}
\end{subequations}
We recall that the geometric data above define the geometry in a coordinate frame which is not explicitly flat at asymptotic infinity. To go to an explicitly asymptotically flat frame one has to apply the transformation rule (\ref{shiftaction}).\footnote{In (\ref{shiftaction}) we give the transformation rule for $a_1$ and $\gamma_2$, but not for $\Theta$. To apply this rule one thus has to derive from the $\tilde \Theta$ in Eq. (\ref{Thetatilde}) the corresponding $\tilde a_1$ and $\tilde \gamma_2$. A possible gauge choice, consistent with the boundary conditions (\ref{tildedasymptotics}),  is
\be
\tilde a_1 = \frac{\tilde K}{V}(d\tau+A)\,,\quad \tilde \gamma_2 =  -Q_5\,\frac{(r^2+a^2)\,\cos^2\theta}{r^2+a^2\cos^2\theta}\,d\phi\wedge d\psi + \delta \tilde \gamma_2\,,
\ee
where
\be
\frac{d}{d v}\delta \tilde \gamma_2= \tilde \Theta - \cald \tilde a_1\,.
\ee
} This geometry is by construction regular and horizon-free. In the next sections we will compute its asymptotic charges and compare with the expectation values of the corresponding operators in the conjectured CFT dual state.

\section{Charges from the dual CFT}\label{section:CFTcharges}

In this section we consider the dual CFT description of the microstate geometries derived in Section~\ref{section:examples}. We use for the CFT the free field formulation summarised in Section~\ref{dctpoint}, even if in this case, the CFT description is far from the gravitational regime. The standard expectation is that the eigenvalues of operators describing conserved charges are not modified when we move away from the orbifold point in the CFT moduli space, nor when we couple the asymptotically flat region. Things are slightly more complicated in our case, because the semiclassical states we are considering are in general a linear combination of different eigenstates of the momentum and angular momentum operators. In the microscopic description this linear combination is determined by the continuous parameter defining the coherent state and of course also on the bulk side the corresponding solution depends on a continuous parameter. The dictionary between these two descriptions is unambiguous only in the decoupling limit, but this relation can have non-trivial corrections that vanish when $R_{\rm AdS}/R \to 0$. In the first example we consider this possibility is not realised and the relation obtained in the decoupling limit can be used also in the asymptotically flat region. On the contrary, the second example discussed in Section~\ref{svd} requires a non-trivial dictionary between the continuous parameters defining the semiclassical microstate and the one appearing in the supergravity solutions. 

\subsection{The dual description of Section~\ref{gvind}}\label{svind}

The first class of solutions we presented is related to descendants $|\psi_\chi\rangle$ of the 2-charge state $|\psi_0\rangle$ in~\eqref{eq:Esk}. We can derive the precise form of the operator connecting these two states by mirroring on the CFT side the gravity construction. We use a spectral flow to bring $|\psi_0\rangle$ back to the NS sector, act with the generator $(J^2_{NS})_0$ and then flow back to the R sector. By using Eq.~\eqref{eq:NS2Rcft}, we can derive the corresponding action in the R sector
\begin{equation}
  \label{eq:J20toR}
  (J^2_{NS})_0 = \frac{1}{2i} \left[(J^+_{NS})_0-(J^-_{NS})_0 \right] 
  \to {\cal J} \equiv \frac{1}{2i} \left[(J^+_{R})_{-1}-(J^-_{R})_{1} \right]~.
\end{equation}
Thus the solution in Section~\ref{gvind}, or better its generalisation discussed in Appendix~\ref{appa}, should correspond to the state
\begin{equation}
  \label{eq:psichi}
  |\psi_\chi\rangle = e^{i \chi^{(2)}_R {\cal J}} |\psi_0 \rangle = 
  e^{\chi \left[(J^+_{R})_{-1}-(J^-_{R})_1 \right]} |\psi_0 \rangle~,
\end{equation}
where in the second step we introduced 
\begin{equation}
  \label{eq:chiRchi} 
\chi\equiv \frac{\chi^{(2)}_R}{2} ~.
\end{equation}
Then for our purposes we can focus on the three generators $(J^+_{R})_{-1}$, $(J^3_{R})_{0}-n_1 n_5/2$, and $(J^-_{R})_{1}$ which satisfy the standard $SU(2)$ algebra
\begin{equation}
  \label{eq:su2a}
  \left[J_0^3 - \frac{n_1n_5}{2} , J^\pm_{\mp 1}\right] = \pm J^\pm_{\mp 1}~,~~~
  \left[ J^+_{-1} , J^-_{1}\right] = 2 \left(J_0^3 - \frac{n_1 n_5}{2}\right)~,
\end{equation}
where from now on we neglect the subscript $R$. In terms of the orbifold free field description we have
\begin{equation}
  \label{eq:Jpmmod}
   J^+_{-1} = i\sum_{\ell} (-\chi_{-1 \ell}^2 \chi_{0 \ell}^1 + \chi_{-1 \ell}^1 \chi_{0 \ell}^2 )~,~~~
   J^-_{1} = i\sum_{\ell} (\bar\chi_{0 \ell}^1 \bar\chi_{1 \ell}^2 - \bar\chi_{0 \ell}^2 \bar\chi_{1 \ell}^1 )~,
\end{equation}
where we explicitely implemented the normal ordering prescription which was understood in the expression of Section~\ref{dctpoint}. Notice that $J^-_{1}$ commutes with the operator $O$ introduced in~\eqref{eq:Osk} and, as expected, annihilates the highest weight RR ground state $|n_1 n_5/2\rangle$. Then, by using\footnote{Maybe the easiest way to derive this identity is to complexify the $SU(2)$ generators so as to obtain the $SL(2,R)$ algebra; then one can realize each exponential as a projective transformation on the complex plane and check that both sides of~\eqref{eq:splitexp} define the same transformation.}
\begin{equation}
  \label{eq:splitexp}
  e^{\chi \left(J^+_{-1}- J^-_1 \right) } = e^{\tan\chi J^+_{-1}}\, (\cos\chi)^{n_1 n_5-2 J^3_0} \, e^{-\tan\chi J^-_{1}}~,
\end{equation}
we can write
\begin{equation}
  \label{eq:JO}
  |\psi_\chi\rangle = \sum_{k} (\cos\chi)^{k} C_k e^{\tan\chi J^+_{-1}}\,  O^k |n_1 n_5/2\rangle~.
\end{equation}
The norm of $\psi_\chi$ is still one since it is obtained from $\psi_0$ by acting with a unitary operator. This can be explicitly checked on the expression above by recalling that $J^-_1$ annihilates $O^k |n_1 n_5/2\rangle$ and by using the identity 
\begin{equation}
  \label{eq:Jspl2}
   e^{\gamma J^-_{1}} e^{\alpha J^+_{-1}} = e^{\frac{\alpha}{1+\alpha\gamma} J^+_{-1}} e^{2\ln(1+\alpha\gamma) \left(\frac{n_1n_5}{2}-J_0^3\right)}  e^{\frac{\gamma}{1+\alpha\gamma} J^-_1}
\end{equation}
with $\alpha=\gamma=\tan\chi$.
At a first sight the results above break down when $\chi\to\pi/2$; let us show that this is not the case. From the explicit expressions~\eqref{eq:Jpmmod}, we see that each term $(J^+_{-1})_\ell$ in $J^+_{-1}$ vanishes when acting on a strand that has eigenvalue $1/2$ for $(J^3_0)_\ell$; also $(J^+_{-1})_\ell$ cannot act twice on the other type of strands present in $\psi_0$ as $(J^+_{-1})_\ell^2 O_\ell |n_1 n_5/2\rangle=0$. Thus the exponential in~\eqref{eq:JO} can always be truncated to a finite sum, but when $\chi\to\pi/2$ only the term $\sim\tan^k\chi$ in this sum can contribute because $\cos\chi\to 0$. Thus we have
\begin{equation}
  \label{eq:Jpi2}
  |\psi_{\frac{\pi}{2}}\rangle = \sum_{k=0}^{n_1n_5} \frac{C_k}{k!} (J^+_{-1})^k\,  O^k |n_1 n_5/2\rangle = \sum_{k=0}^{n_1n_5} C_k \,  \hat{O}^k |n_1 n_5/2\rangle ~,
\end{equation} 
where $\hat{O}=\sum_\ell (J^+_{-1})_\ell O_\ell$. We see that $\chi=\pi/2$ represents the limiting case in which the operator $J^+_{-1}$ has acted once on all available strands.\footnote{We thank S. Mathur for drawing our attention to this point.} For this value of $\chi$, we obtain a state that is similar to the original 2-charge state $|\psi_0\rangle$, but now each action operator $\hat{O}$ on the highest weight state increases the eigenvalues of $J^3_0$ and $L_0$ by $1/2$ and $1$ respectively, while it still decreases the eigenvalue of $\tilde{J}^3_0$ by $1/2$.

We have now written $|\psi_\chi\rangle$ in a form that makes it easy to calculate, for any $\chi$, the average values of the momentum operator $L_0-\tilde{L}_0$ and the angular momenta $J_0^3$ and $\tilde{J}_0^3$. Of course, since the operator $O$ is holomorphic, the expectation values of tilded operators in the state $|\psi_\chi\rangle$ are independent of $\chi$ and will be identical to the ones obtained in the state $|\psi_0\rangle$
\begin{equation}
  \label{eq:vevpsi0}
  \langle \psi_0| \tilde{L}_0 | \psi_0\rangle = 0~,~~~
  \langle \psi_0| \tilde{J}^3_0 | \psi_0\rangle = \frac{1}{2} \sum_{k} (n_1 n_5-k) C_k^2  = \frac{n_1 n_5}{2} A^2 = \frac{n_1 n_5}{2} \frac{1}{1+\eta^2}~,
\end{equation}
where $C_k$ is defined in~\eqref{eq:Ck} and in the last identity we used~\eqref{AandB}.

The expectation values of $L_0$ and $J_0^3$ are identical to~\eqref{eq:vevpsi0} at zero order in $\chi$. In general both of them receive the same $\chi$-dependent correction ${\mathcal C}_\chi$ as the commutation relations of $J^+_{-1}$ with $J_0^3$ are identical to those with $L_0$
\begin{equation}
  \label{eq:vevdchi}
  \langle \psi_\chi| L_0 | \psi_\chi\rangle = {\mathcal C}_\chi~,~~~
  \langle \psi_\chi| J^3_0 | \psi_\chi\rangle = \frac{n_1 n_5}{2} \frac{1}{1+\eta^2} + {\mathcal C}_\chi~.
\end{equation}
In order to derive ${\mathcal C}_\chi$ it is convenient to realise $J^+_{-1}$ as a multiplicative operator ($J^+_{-1} \to \xi$) and $J_0^3$ and $L_0$ as differential operators ($J_0^3\,,~L_0 \to\xi \partial_\xi$):
\begin{equation}
  \label{eq:Ckd1}
  [L_0,J^+_{-1}] = J^+_{-1}~,~~ [J_0^3,J^+_{-1}] = J^+_{-1} ~~ \to ~~ [\xi \partial_{\xi},{\xi}] = {\xi}~.
\end{equation}
Then we can write the commutator $[L_0,e^{\alpha J^+_{-1}}]$ as $\alpha\, \partial_\alpha e^{\alpha J^+_{-1}}$ and use~\eqref{eq:Jspl2} to obtain
\begin{align}
  {\mathcal C}_\chi & = \sum_{h,k} C_h C_k (\cos\chi)^{h+k} \langle n_1 n_5/2| O^{\dagger h} e^{\tan\chi J^-_1} [L_0,e^{\tan\chi J^+_{-1}}] O^k |n_1 n_5/2\rangle \\ \nonumber
 & = \sum_{k} C_k^2 (\cos\chi)^{2k} \alpha\, \partial_\alpha e^{k\ln(1+\alpha\gamma)} \Big|_{\stackrel{\scriptstyle \alpha=\tan\chi}{\scriptstyle \gamma=\tan\chi} } = \sin^2\chi \sum_k k\, C_k^2 =  n_1 n_5 \frac{\eta^2\sin^2\chi}{1+\eta^2}~,
\end{align}
where in the final step one can use the result of Eq.~\eqref{eq:vevpsi0} for the sum over $k$.

Summarizing, we have
\begin{subequations}
  \label{eq:av1}
  \begin{align}
  \langle n_p \rangle& = \langle \psi_\chi| L_0 |\psi_\chi\rangle = n_1 n_5  \frac{\eta^2\sin^2\chi}{1+\eta^2}~, \\
  \langle J \rangle &= \langle \psi_\chi| J^3_0 |\psi_\chi\rangle = \frac{n_1 n_5}{2} \frac{1}{1+\eta^2} + n_1 n_5 \frac{\eta^2\sin^2\chi}{1+\eta^2}~, \\
  \langle \tilde{J} \rangle & = \langle \psi_\chi| \tilde{J}^3_0 |\psi_\chi\rangle =  \frac{n_1 n_5}{2} \frac{1}{1+\eta^2} ~.
  \end{align}
\end{subequations}

\subsection{The dual description of Section~\ref{gvd}} \label{svd}

Let us now turn to the second class of solutions we discussed in some detail in Section~\ref{gvd}. 
The change of coordinates (\ref{J3minusn}) we apply in the decoupling limit of the 2-charge solution 
(\ref{luninmathurtwocharge}) is identified with the CFT operator $J^3_{-n} - J^3_{n}$.
Hence the CFT state dual to the solution (\ref{solutiondatafinal}) should be\footnote{All the CFT operators in the section are in the R sector.}
\begin{equation}
  \label{eq:J3nd}
  |\phi_\epsilon\rangle = e^{\epsilon\, (J^3_{-n} - J^3_{n})}\, |n_1 n_5/2\rangle~,
\end{equation}
where $\epsilon = \hat \epsilon$ and we choose $n$ to be positive. Note that the identification between $\epsilon$ and $\hat \epsilon$ is justified only in the decoupling limit, and indeed we will see in the next Section that the two parameters differ after extending the geometry to the asymptotically flat region. 

Since $J_0^3$ commutes with all modes $J^3_m$, then it is clear that $ |\phi_\epsilon\rangle$ is still an eigenstate of both $J^3_0$ and $\tilde{J}^3_0$ with the same eigenvalues of the 2-charge highest weight state $|n_1 n_5/2\rangle$
\begin{equation}
  \label{eq:Jv2}
   \langle J \rangle =  \langle \phi_\epsilon| J^3_0 |\phi_\epsilon \rangle = \frac{n_1 n_5}{2}~,~~~ \langle \tilde{J} \rangle = \langle \phi_\epsilon| \tilde{J}^3_0 |\phi_\epsilon \rangle = \frac{n_1 n_5}{2}~.
\end{equation}

In order to calculate the average value for the momentum operator we can proceed as done above for the state $|\psi_\chi\rangle$. We first separate the destruction and the creation operators in the exponent
\begin{equation}
  \label{eq:phiese}
  |\phi_\epsilon\rangle = e^{-\frac{n_1 n_5}{2} \frac{n \epsilon^2}{2}} e^{\epsilon J^3_{-n}} \,|n_1 n_5/2\rangle~
\end{equation}
and then write the average momentum as
\begin{equation}
  \label{eq:avm2}
  \langle n_p \rangle = \langle \phi_\epsilon| L_0 |\phi_\epsilon \rangle = e^{-(n_1 n_5) \frac{n \epsilon^2}{2}} \langle n_1 n_5/2 |\, e^{\epsilon J^3_{n}} [L_0,  e^{\epsilon J^3_{-n}}] \, |n_1 n_5/2\rangle~.
\end{equation}
Again we can realise $J^3_{-n}$ as a multiplicative operator ($J^3_{-n}\to {\xi}$ ) and $L_0$ as a differential one ($L_0 \to n\, {\xi}\partial_{\xi}$). Finally we obtain
\begin{equation}
  \label{eq:np2}
  \langle n_p\rangle =  e^{-(n_1 n_5) \frac{n \epsilon^2}{2}}  \left[n\,\alpha\, \partial_\alpha e^{(n_1 n_5) \frac{n \alpha\gamma}{2}}\right]_{\stackrel{\scriptstyle \alpha=\epsilon}{\scriptstyle \gamma=\epsilon} }  = \frac{n_1 n_5}{2} \epsilon^2 n^2~.
\end{equation}

\section{Matching CFT and gravity}
\label{section:CFTmatching}
The generating technique we used to construct the geometries in Sections~\ref{gvind} and \ref{gvd} allows us to identify uniquely the CFT states dual to the geometries in the decoupling limit. We started from 2-charge solutions dual to known RR ground states and acted on the near-horizon region of these solutions by change of coordinates that realize the chiral algebra of the CFT: hence the asymptotically AdS solutions constructed in this way are dual to specific  CFT superdescendants. The solution in Eqs.~(\ref{nhvind}) is dual to the state (\ref{eq:psichi}) and the solution in (\ref{nearhorizonJ3}) is dual to (\ref{eq:J3nd}). The extension of the near-horizon
geometries to the asymptotically flat region is uniquely determined by regularity and asymptotic flatness conditions, at least within our ansatz: one thus obtains the solutions (\ref{3chargefin}) and (\ref{solutiondatafinal}). In the asymptotically flat geometries, however, the identification of the continuos parameters characterizing the microstates on the gravity and on the CFT sides is a priori not obvious.

In this Section we compute the angular momenta and momentum derived from the asymptotic region of the geometries and compare them with the expectation values in the dual CFT states of the R-charges $J^3_0$ and $\tilde J^3_0$ and of the momentum operator $L_0-\tilde L_0$.  This comparison will provide both non-trivial checks of the duality between states and geometries and also the relation between the gravity and the CFT parameters.

\subsection{Asymptotic charges from gravity}
Let us review the well known procedure to extract the asymptotic charges from the geometry \cite{Myers:1986un}. One should work in a coordinate frame that satisfies the harmonic gauge condition
$\partial_\mu (\sqrt{|g| g^{\mu\nu}})=0$ at large distances: if one expands the metric at linear order around the flat asymptotic background $M^{1,4}\times S^1\times T^4$, the harmonic gauge is satisfied if $ds^2_4$ is flat up to corrections of order $O(r^{-3})$ and if the following conditions are satisfied
\be\label{deDonder}
d *_4 \beta=0\,,\quad \partial_v \beta =0\,,\quad d *_4 \omega-2\, \partial_v Z=0 
\ee
at leading order in $1/r$, where $\beta$ and $\omega$ vanish at infinity like $1/r^3$ and $Z=1+O(r^{-2})$.

In such a coordinate system one has the large $r$ expansions
\be\label{asymptoticcharges}
Z_1\approx 1+\frac{Q_1}{r^2}\,,\,\, Z_2\approx 1+\frac{Q_5}{r^2}\,,\,\, -\frac{\mathcal{F}}{2}\approx \frac{Q_p}{r^2}\,,\,\, \frac{\beta_\phi+\omega_\phi}{\sqrt{2}}\approx \frac{J_\phi}{r^2}\sin^2\theta\,,\,\, \frac{\beta_\psi+\omega_\psi}{\sqrt{2}}\approx \frac{J_\psi}{r^2}\cos^2\theta\,.
\ee
In principle the coefficients of the $1/r$ expansion could depend on $v$, but we have kept only the $v$ zero-modes, because they are the only ones contributing to the global charges. 

The gravity coefficients are quantized in the following way
\be\label{quantizedcharges}
Q_1 = \frac{(2\pi)^4\,g_s\,\alpha'^3}{V_4} n_1\,,\quad Q_5 = g_s\,\alpha' n_5\,,\quad Q_p = \frac{(2\pi)^4\,g_s^2 \alpha'^4}{V_4\,R^2}\,n_p\,,\quad J_{\phi,\psi}= \frac{(2\pi)^4\,g_s^2 \alpha'^4}{V_4\,R}\,j_{\phi,\psi}\,,
\ee
where $g_s$ is the string coupling, $V_4$ is the volume of $T^4$ and $R$ is the radius of the $S^1$; $n_1$ and $n_5$ count the numbers of D1 and D5 branes, $n_p/R$ is the momentum along $S^1$ and $j_\phi$ and $j_\psi$ are the two angular momenta of $\mathbb{R}^4$. 

Let us now apply this recipe to compute the asymptotic charges of the solution (\ref{3chargefin}). One can check that the harmonic gauge conditions (\ref{deDonder}) are satisfied, and one can thus directly read off the large $r$ coefficients (\ref{asymptoticcharges}) from the geometry:
\be
Q_p = \frac{b^2}{2}\,,\quad J_\phi=R\Bigl(a^2+\frac{b^2}{2}\Bigr)\,,\quad J_\psi = \frac{R\,b^2}{2}\,.
\ee
Using the radius relation (\ref{Q1Q5Rab}) and the definition (\ref{etadef}), the quantized charges defined in (\ref{quantizedcharges}) are
\be
n_p = n_1 n_5\,\frac{\eta^2}{1+\eta^2}\,,\quad j_\phi =n_1 n_5\,,\quad j_\psi=n_1 n_5\,\frac{\eta^2}{1+\eta^2}\,.
\ee
For comparison with the CFT it is useful to consider the extension of the solution (\ref{Q1Q5Rab})
to generic values of the rotation parameter  $\chi$ appearing in Eq. (\ref{eq:chiRchi}). The derivation of the metric is given in the Appendix; we quote here the corresponding quantized charges
\be\label{chargesGEN}
n_p = n_1 n_5\,\frac{\sin^2\!\chi\, \eta^2}{1+\eta^2}\,,\quad j_\phi =n_1 n_5\frac{1+\sin^2\!\chi\, \eta^2}{1+\eta^2}\,,\quad j_\psi=n_1 n_5\,\frac{\sin^2\!\chi \,\eta^2}{1+\eta^2}\,.
\ee
To compare with the CFT result (\ref{eq:av1}) we have to relate $j_\phi$ and $j_\phi$ with the right-moving and left-moving CFT quantum numbers $j$ and $\tilde j$: from (\ref{J3}) and (\ref{tildeJ3}) we find
\be
j = \frac{j_\phi+j_\psi}{2}\,,\quad \tilde j = \frac{j_\phi-j_\psi}{2}\,.
\ee
We thus see that the gravity charges exactly match the averages of the corresponding CFT operators in the dual microstate: in this case the rotation parameter $\chi$ of the gravity solution is directly identified with the parameter that appears in the state (\ref{eq:psichi}).

The computation of the asymptotic charges for the solution (\ref{solutiondatafinal}) is a bit more involved. The metric does not have the large $r$ behavior expected in the harmonic gauge: 
the $d t \,d x^i$ terms, associated with the 1-form $\hat\omega$, have a $1/r$ fall-off, instead of the expected $1/r^3$: 
\be
\hat\omega\approx -\frac{n^2 \hat\epsilon\,a^2}{\sqrt{2}\,R}\,\sin\Bigl(\frac{\sqrt{2}\, n\,v}{R}\Bigr)\Bigl(\frac{\cos2\theta}{r} d r -\sin2\theta\,d\theta \Bigr)\,.
\ee
Moreover the 4D metric $d\hat s^2_4$ has non-trivial corrections of order $1/r^2$:
\be
d\hat s^2_4\approx dx^i dx^i -\beta_k \dot{f}^k \,dx^i dx^i + (\beta_j \,\dot{f}_i+\beta_i \,\dot{f}_j)\,dx^i dx^j \,.
\ee
The unwanted terms in both $\hat \omega$ and $d\hat s^2_4$ can be reabsorbed by the change of variables\footnote{The metric after this change of coordinates still does not obey the harmonic gauge condition, because the last relation in (\ref{deDonder}) is not fulfilled. One can however restore the gauge condition by a coordinate transformation of the form $u\to u + U(r,\theta,v)$, for some function
$U(r,\theta,v)$ whose $v$-integral vanishes. Hence this further change of coordinates has no influence on the global charges, and we will ignore it in the following.}
\be
r\to r +n\,\hat\epsilon\,a^2 \cos\Bigl(\frac{\sqrt{2}\, n\,v}{R}\Bigr)\, \frac{\cos2\theta}{2\,r}\,,\quad \theta\to \theta- n\,\hat\epsilon\,a^2 \cos\Bigl(\frac{\sqrt{2}\, n\,v}{R}\Bigr)\, \frac{\sin2\theta}{2\,r^2}\,.
\ee
Expanding the metric at large $r$ after this change of variables one finds
\be
Q_p=\frac{\hat\epsilon^2 n^2}{4\,R^2}(2\,Q_1 Q_5+2\,n\,a^2 (Q_1+Q_5)+n^2 a^4) \,,\quad J_\phi= \frac{Q_1 Q_5}{R}\,,\quad J_\psi=0\,,
\ee
and thus the quantized charges are
\be\label{npfull}
n_p=\,n_1 n_5\,\hat\epsilon^2 n^2 \,\frac{2\,Q_1 Q_5+2\,n\,a^2 (Q_1+Q_5)+n^2 a^4}{4\,Q_1 Q_5}\,,\quad j=\frac{n_1 n_5}{2}\,,\quad \tilde j =\frac{n_1 n_5}{2} \,.
\ee
Comparison of these results with the CFT predictions (\ref{eq:Jv2}) and (\ref{eq:np2}) shows an agreement at the level of the angular momenta: despite the fact that the angular momenta are $\epsilon$-independent, this agreement is a non-trivial consequence of the regularity requirement of 
the gravity solution. The comparison of the momentum charge $n_p$ provides the relation between the CFT parameter $\epsilon$ and the gravity one $\hat \epsilon$:
\be\label{epsilonrelation}
\epsilon^2 = \hat\epsilon^2 \,\frac{2\,Q_1 Q_5+2\,n\,a^2 (Q_1+Q_5)+n^2 a^4}{2\,Q_1 Q_5}\,.
\ee
In the decoupling limit, in which $a^2\ll Q_1, Q_5$, one recovers the result $\epsilon=\hat\epsilon$, as it is required by the identification of the near-horizon geometry and the CFT state.
\section{Discussion}
\label{section:discussion}

In this paper we studied a class of 3-charge configurations in the D1-D5-P system both from the gravity and the dual CFT point of view. The states we analysed are certainly very particular: they correspond to superdescendants of a small class of 2-charge states. Their simplicity allows for an analytic treatment: on the bulk side we could derive explicit solutions of all supergravity equations and check that they were regular in the interior, while on the CFT side we used the free field description at the orbifold point. Even if the states we studied are special, they have some new interesting features with respect to the known solutions and so might come closer to capturing the behaviour of generic configurations.

A first basic property of the states considered in this paper is that
they are not eigenvectors of the momentum operator. On the CFT side
this means that they are a linear combinations of terms with different
momentum eigenvalues, as it can be seen by Taylor expanding the
exponential in~\eqref{eq:psichi} and~\eqref{eq:J3nd}. One can follow
the discussion of the end of Section~\ref{dctpoint} and characterise
our 3-charge solutions as semiclassical configurations with an average
value for the momentum, but also a width including many states with
different eigenvalues. On the gravity side this is reflected by the
fact that $\partial_v$ is not a killing vector: the solutions we
presented, see Eqs.~\eqref{3chargefin} and~\eqref{solutiondatafinal},
depend on $v$ explicitly. The $v$-independent solutions derived
in~\cite{Giusto:2004id,Giusto:2004ip} correspond to particular
momentum eigenvectors on the dual CFT side and presumably the same
should hold for the generalisation discussed
in~\cite{Bena:2005va,Berglund:2005vb} even if in these cases the
precise dual states are not known.

In the decoupling limit the expansion of the supergravity solutions in
the asymptotically AdS region is related to the expectation values of
certain $1/2$-BPS operators of the dual CFT in the corresponding
states. This relation was discussed quantitatively
in~\cite{Kanitscheider:2006zf,Kanitscheider:2007wq} for the 2-charge
case as the relevant expectation values are protected by supersymmetry
and the results from the orbifold CFT and the dual gravity description
should and do match. In the 3-charge case studied in this paper a
similar quantitative agreement is not expected, but the generic
qualitative features of the CFT correlators can still be reproduced by
the supergravity solution. For instance a vacuum expectation value in
an eigenstate of the momentum operator will be non-trivial only for
CFT operators that carry zero momentum, which corresponds to a
$v$-independent geometry at least in the asymptotically AdS region. On
the contrary states that are a superposition of different momentum
eigenstates can excite operators with a non-trivial Kaluza-Klein mode
along the $S^1$ corresponding to $v$-dependent geometries.\footnote{We
thank K. Skenderis and M. Taylor for an enlightening discussion on
this point.}

Another interesting feature that can appear in the class of solutions we considered is displayed in the example of Section~\ref{gvd}. This configuration is rather complicated when expressed in the coordinates that are appropriate to read the charges. This is true both in the asymptotically flat region, where we need to use the hatted quantities of Eqs.~\eqref{shiftaction} together with the change of variables discussed in Section~\ref{section:CFTmatching}, and in the near horizon region, where the hatted quantities of Eqs.~\eqref{nearhorizonJ3} are the appropriate ones. From this point of view, this example shares the same property of the asymptotically AdS configurations recently discussed in~\cite{Shigemori:2013lta}. In our case we could also extend the solution to the asymptotically flat region and, as discussed in Section~\ref{section:CFTmatching}, this completion shows an unexpected feature: if we insist to keep the identification obtained in the decoupling limit between the parameter defining the coherent state and the one appearing in the supergravity solution, then the average value momentum derived from supergravity does depend on the moduli at infinity (hidden in the ratio $a^2/\sqrt{Q_1 Q_5}$). In the large charge limit (where $a^2/\sqrt{Q_1 Q_5}\to 0$) the supergravity and the microscopic results match straightforwardly; however, as we expect that this match holds also for the asymptotically flat solutions (i.e. for any value of $R$), we proposed the identification~\eqref{epsilonrelation}. Of course the eigenvalues of conserved operators do not depend on this redefinition and so we can read from the geometry the values of the angular momenta $j$'s in a straightfoward way, see~\eqref{npfull}. Notice that, even if in the dual CFT description the $j$'s are obviously $\epsilon$-independent, on the gravity side this is the result of a non-trivial cancellation, which supports our identification between the microstate considered and the geometry~\eqref{solutiondatafinal}. 

We argue that identifications such as the one in~\eqref{epsilonrelation} do not represent a contradiction but possibly are common to a large class of 3-charge semiclassical configurations since the definition of both the $v$-dependent geometries and their dual states depends on a continuous parameter. Also it may not be surprising that a similar phenomenon does not appear in the 2-charge case, since in that case one considers $1/4$ rather than $1/8$-BPS configurations and the higher amount of supersymmetry puts more stringent constraints. Another possibility would be to interpret the solution in~\eqref{solutiondatafinal} as an unbound state where there is another momentum carrying perturbation in the asymptotically flat region that is independent from the one we started with in~\eqref{eq:J3nd}. We find this interpretation less attractive, as usually unbound systems are not dual to regular geometry. Also it would be necessary to assume that there is another smooth solution that has the same decoupling limit as the one discussed in Section~\ref{gvd}; while we cannot exclude this possibility in general, this requires to go beyond the ansatz we considered in this paper, for instance by relaxing the assumptions on $\beta$. Of course it would be interesting to study this issue in more detail and possibly to provide further evidence supporting the interpretation proposed here.

Let us conclude by some brief comments on a possible generalisation of our approach. The two examples we discussed inherited several features from the parent 2-charge geometry. For instance there exists a coordinate system, where the base metric $ds_4^2$ in~\eqref{ansatzsummary} is Euclidean. However, it should not be difficult to consider cases whose base geometry is a two centre GH space. For instance one could follow the approach of~\cite{Giusto:2004id,Giusto:2004ip} and take the spectral flow (on the left sector only) of the simpler example discussed in Section~\ref{gvind}. This should yield a new geometry, with a more complicated base, but still falling in the class studied in this paper and for which one could hope to find an extension to the asymptotic flat region. In general one can probably find a new class of multi-centre geometries by following~\cite{Bena:2005va,Berglund:2005vb} that however are now $v$-dependent. 

\vspace{7mm}
 \noindent {\large \textbf{Acknowledgements} }

 \vspace{5mm} 

We would like to thank I. Bena, O. Lunin, L. Martucci, S. Mathur, M. Petrini, M. Shigemori, K. Skenderis, M. Taylor, D. Turton, and N. Warner for useful discussions and correspondence.
This research is partially supported by CNRS, by STFC (Grant ST/J000469/1, {\it String theory, gauge 
 theory \& duality}), by the MIUR- PRIN contract 2009-KHZKRX, by the Padova University Project CPDA119349 and by INFN. R.R. wishes to thank the {\it Institut Lagrange de Paris} for hospitality 
and support during the completion of this work.

\appendix
\section{Solution for generic rotation}\label{appa}
In this appendix we work out the generalization of the solution (\ref{3chargefin}) for a generic rotation parameter $\chi\equiv \frac{\chi^{(2)}_R}{2}$. 

We begin by rewriting the near-horizon solution obtained after the rotation (\ref{rotg}) with a generic parameter $\chi$, in the formalism of Section~\ref{section:Warner}; we will see that one needs a slight generalization of that formalism. The near-horizon expression of the function $Z_4$ is
\be
Z_4 =  \frac{R\, a\, b}{\sqrt{r^2+a^2}\,(r^2+a^2 \cos^2\theta)}\,\Bigl[\cos\chi \,\sin\theta\,\cos\phi + \sin\chi\,\cos\theta\,\cos\hat v\Bigr]\,.
\ee
Comparing with Eqs. (\ref{Z4harm}), (\ref{K4L4}) one derives the generalized harmonic functions $K_4$ and $L_4$
\be
K_4 = 2\sqrt{2}\,a\, b\, \sin\chi\,\frac{\cos\theta}{\sqrt{r^2+a^2}\,(r^2+a^2\sin^2\theta)}\,\cos\hat v\,,
\ee
\be
L_4 = \frac{R\,a\, b}{\sqrt{r^2+a^2}}\Bigl[\cos\chi\,\frac{\sin\theta}{r^2+a^2\cos^2\theta}\,\cos\phi + \sin\chi\,\frac{\cos\theta}{r^2+a^2\sin^2\theta}\,\cos\hat v\Bigr]\,.
\ee
Analogously, from the form of the near-horizon expression for $Z_2$
\be
Z_2 = \frac{Q_5}{r^2+a^2\sin^2\theta}\,,
\ee
we find
\be
K_1=0\,,\quad L_2 = \frac{Q_5}{r^2+a^2\sin^2\theta}\,.
\ee
For $\mathcal{F}$ we have
\be\label{eq:a6}
\mathcal{F}=-\sin^2\chi\,\frac{b^2}{r^2+a^2}\,.
\ee
From Eqs.~\eqref{eq:a6} and (\ref{calFharm}) it follows that
\be
L_3=-\frac{b^2\,\sin^2\chi}{r^2+a^2\sin^2\theta}\,\Bigl[1+\frac{a^2\,\cos^2\theta}{r^2+a^2}\,\cos2\hat v\Bigr]\,.
\ee
The situation becomes more complicated when one looks at $Z_1$:
\begin{align}
Z_1 &=  \frac{R^2}{Q_5} \frac{a^2+\frac{b^2}{2}}{r^2+a^2 \cos^2\theta}+\frac{R^2\, a^2\, b^2}{2\,Q_5}\,\frac{1}{(r^2+a^2 \cos^2\theta)(r^2+a^2)}\,\Bigl[\cos^2\chi\,\sin^2\theta\,\cos2\phi \nonumber\\
&\qquad+\sin\chi \cos\chi\,\sin2\theta\,\cos(\hat v + \phi) + \sin^2\chi\,\cos^2\theta \,\cos 2 \hat v \Bigr]\,.
\end{align}
One can see that this $Z_1$ cannot be rewritten in the form (\ref{Z1harm}) with some generalized harmonic functions $K_2$ and $L_1$ satisfying (\ref{K2L1}); the problematic term is the one proportional to $\cos(\hat v+\phi)$, which vanishes at both $\chi=0$ and $\chi=\frac{\pi}{2}$. To deal with this term one needs to allow a more general form for the flux $\Theta_2$ than the one considered in (\ref{Theta2harm}). The generalization\cite{Niehoff:2013kia} requires the introduction of a 1-form $\lambda_2$ with no components along $d\tau$, so that $\Theta_2$ can be written as
\be
\Theta_2 = \Bigl[\mathcal{D}\Bigl(\frac{K_2}{V}\Bigr)+\lambda_2\Bigr]\wedge (d\tau+A) + *_4 \Bigl\{\Bigl[\mathcal{D}\Bigl(\frac{K_2}{V}\Bigr)+\lambda_2\Bigr]\wedge (d\tau+A) \Bigr\}\,.
\ee
There is clearly a large arbitrariness in the choice of $K_2$ and $\lambda_2$, that one can exploit to
impose the usual constraint
\be
\partial_\tau K_2 + \partial_v L_1=0\,.
\ee
With this choice it is easy to extract $K_2$ and $L_1$ from $Z_1$:
\be\label{K2gen}
K_2 = \frac{\sqrt{2}\, b^2 R}{Q_5\,(r^2+a^2)}\,\Bigl[\sin^2\chi\, \frac{a^2\,\cos^2\theta}{(r^2+a^2\sin^2\theta)}\,\cos 2\hat v - \sin\chi\cos\chi\,\tan2\theta\,\cos(\hat v + \phi)\Bigr]\,,
\ee
\begin{align}
L_1 =& \frac{R^2}{Q_5} \frac{a^2+\frac{b^2}{2}}{r^2+a^2 \cos^2\theta}\\
&+\frac{R^2\, a^2\, b^2}{2\,Q_5\,(r^2+a^2)}\,\Bigl[\cos^2\chi\,\frac{\sin^2\theta}{r^2+a^2\cos^2\theta}\,\cos2\phi+\sin^2\chi\,\frac{\cos^2\theta}{r^2+a^2\sin^2\theta}\,\cos 2\hat v\Bigr]\,. \nonumber
\end{align}
Moreover, comparing with the near-horizon expression for $\Theta_2$, one derives
\begin{align}
\lambda_2 &= \frac{b^2 R}{2\sqrt{2}\,Q_5\,(r^2+a^2)}\,\sin\chi\cos\chi\,\Bigl[
r^2 \tan2\theta \,(d\psi-d\phi)\,\sin(\hat v + \phi) \nonumber\\ & + 
\Bigl(\frac{a^2 \,r\,\tan2\theta}{r^2+a^2}\,dr + 2 \frac{r^2+2 a^2 \sin^2\theta\cos^2\theta}{\cos^22\theta}\,d\theta\Bigr)\cos(\hat v+\phi)\Bigr]\,.
\end{align}
In this more general setting the generalized harmonicity condition for $K_2$ is deformed to
\be\label{eqK2GEN}
*_4\mathcal{D}*_4\mathcal{D}K_2 + V *_4\mathcal{D}*_4\lambda_2+2 *_4 (dV\wedge *_4 \lambda_2)=0\,,
\ee
and one can check that the $K_2$ in (\ref{K2gen}) satisfies this condition.
The 1-form $\lambda_2$ can be shown to satisfy
\be\label{eqlambda2}
K_3 *_4\mathcal{D}*_4\lambda_2+2 *_4 (dK_3\wedge *_4 \lambda_2)=0\,.
\ee
We are left with the task of finding the function $M$, which determines $\omega$ according to the usual relations (\ref{muharm}), (\ref{zetaeqs}).
$M$ should satisfy the deformed harmonicity condition
\be\label{eqMGEN}
2*_4\mathcal{D}*_4\mathcal{D}M - L_2 *_4\mathcal{D}*_4\lambda_2-2 *_4 (dL_2\wedge *_4 \lambda_2)=0\,.
\ee 
Exploiting the fact that 
\be
L_2 = \frac{Q_5}{\sqrt{2}R}\,K_3 +\frac{Q_5}{4}\,V\,,
\ee
and the relations (\ref{eqK2GEN}), (\ref{eqlambda2}), one sees that a solution for $M$ is
\be
M = -\frac{Q_5}{8}\,K_2 \,.
\ee
So in general one has
\be
 M=\frac{b^2 R}{4\sqrt{2}}\, \sin\chi\cos\chi\,\frac{\tan2\theta}{r^2+a^2}\,\cos(\hat v + \phi)+M_\mathrm{harm}\,,
\ee
where $M_\mathrm{harm}$ is a generalized harmonic function that is determined by regularity. After imposing the regularity constraints described in Section \ref{section:chiralalgebra}, one finds
\begin{align}
M=&\frac{R \,a^2}{2\sqrt{2}(r^2+a^2\cos^2\theta)}+\frac{R\,b^2\,\sin^2\chi}{4\sqrt{2}}\Bigl(\frac{1}{r^2+a^2\cos^2\theta}+\frac{1}{r^2+a^2\sin^2\theta}\Bigr)\\ \nonumber
&+\frac{R\,a^2\,b^2\,\sin^2\chi}{4\sqrt{2}}\frac{\cos^2\theta\cos2\hat v}{(r^2+a^2)(r^2+a^2\sin^2\theta)} + \frac{R\,b^2\,\sin\chi\cos\chi}{4\sqrt{2}} \frac{\tan2\theta}{r^2+a^2}\,\cos(\hat v+\phi)\,.
\end{align}

The extension to the asymptotically flat region proceeds as usual: one adds a ``1'' to $L_1$ and $L_2$ and deforms a coefficient of the functions $K_2$ and $L_1$ in order to have a regular solution:
\be
K_2 = \frac{\sqrt{2}\, b^2 R}{(r^2+a^2)}\,\Bigl[\frac{\sin^2\chi}{Q_5+a^2}\, \frac{a^2\,\cos^2\theta\cos 2\hat v}{(r^2+a^2\sin^2\theta)} - \frac{\sin\chi\cos\chi}{Q_5}\,\tan2\theta\,\cos(\hat v + \phi)\Bigr]\,,
\ee
\begin{align}
L_1 & = 1+\frac{R^2}{Q_5} \frac{a^2+\frac{b^2}{2}}{r^2+a^2 \cos^2\theta}\\ \nonumber
& + \frac{R^2\, a^2\, b^2}{2\,(r^2+a^2)}\,\Bigl[\frac{\cos^2\chi}{Q_5}\,\frac{\sin^2\theta}{r^2+a^2\cos^2\theta}\,\cos2\phi+\frac{\sin^2\chi}{Q_5+a^2}\,\frac{\cos^2\theta}{r^2+a^2\sin^2\theta}\,\cos 2\hat v\Bigr]\,.
\end{align}
Moreover, as a consequence of Eq. (\ref{eqMGEN}), the addition of ``1'' to $L_2$ generates a new contribution to $M$ that is determined by the differential equation
\be
2*_4\mathcal{D}*_4\mathcal{D}\,\delta M - *_4\mathcal{D}*_4\lambda_2=0\,,
\ee
whose solution is
\be
\delta M = \frac{R\,b^2\,\sin\chi\cos\chi}{4\sqrt{2}\,Q_5}\,\frac{r^2+\frac{a^2}{2}}{r^2+a^2}\,\tan2\theta\,\cos(\hat v+\phi)\,.
\ee
Hence, the function $M$ of the asymptotically flat solution is
\begin{align}
M& = \frac{R \,a^2}{2\sqrt{2}(r^2+a^2\cos^2\theta)}+\frac{R\,b^2\,\sin^2\chi}{4\sqrt{2}}\Bigl(\frac{1}{r^2+a^2\cos^2\theta}+\frac{1}{r^2+a^2\sin^2\theta}\Bigr)\nonumber\\
&+\frac{R\,a^2\,b^2\,\sin^2\chi}{4\sqrt{2}}\frac{\cos^2\theta}{(r^2+a^2)(r^2+a^2\sin^2\theta)}\,\cos2\hat v \\ \nonumber
&+ \frac{R\,b^2\,\sin\chi\cos\chi}{4\sqrt{2}} \frac{\tan2\theta}{r^2+a^2}\,\Bigl(1+\frac{r^2+\frac{a^2}{2}}{Q_5}\Bigr)\,\cos(\hat v+\phi)\,.
\end{align}
The remaining functions -- $V$, $K_3$, $K_1$, $K_4$, $L_3$ -- are the same as in the near-horizon.
To have the full solution one should still compute the $\mathbb{R}^3$ part of $\omega$, which requires solving a cumbersome system of partial differential equations. Since our main interest here is to derive the asymptotic charges of the solution for generic $\chi$, and this does not require the knowledge of the full $\omega$, we will not solve this problem here. 

According to the definitions (\ref{asymptoticcharges}), the momentum charge is easily extracted from 
$\mathcal{F}$ 
\be
Q_p = \sin^2\chi\,\frac{b^2}{2}\,.
\ee
The angular momenta can be read off solely from $\mu$. Indeed at large distances, and restricting only to the $v$-independent terms, one has
\begin{align}
\frac{\beta+\omega}{\sqrt{2}} & \approx \frac{J_\psi \cos^2\theta\, d\psi + J_\phi \sin^2\theta \,d\phi}{r^2}\nonumber\\
&=\frac{(J-\tilde J \cos2\theta)\,(d\psi+d\phi) + (J \cos2\theta - \tilde J)\,(d\psi-d\phi)}{2\,r^2}\,,
\end{align}
where
\be
J=\frac{J_\phi+J_\psi}{2}\,,\quad \tilde J = \frac{J_\phi-J_\psi}{2}\,.
\ee
Hence the knowledge of the $d\psi+d\phi$ component of $\omega$, as a function of $\theta$, is enough to derive both angular momenta. We find
\be
J = \frac{R}{2}\,(a^2+\sin^2\chi \,b^2)\,,\quad \tilde J = \frac{R}{2}\,a^2\,.
\ee
Converting to the quantized charges one finds the results reported in (\ref{chargesGEN}).

\providecommand{\href}[2]{#2}\begingroup\raggedright\endgroup

\end{document}